\documentclass[prb,superscriptaddress,twocolumn,showpacs,amsmath,amssymb]
{revtex4}
\usepackage{graphicx}
\usepackage{dcolumn}
\usepackage{bm}
\usepackage{txfonts}
\usepackage{mathrsfs}
\usepackage{feynmf}
\usepackage{comment}
\def\bi{{\mathbf i}}
\def\bj{{\mathbf j}}
\def\bk{{\mathbf k}}

\def\bq{{\mathbf q}}
\def\bA{{\mathbf A}}
\def\bG{{\mathbf G}}
\def\bR{{\mathbf R}}
\def\b0{{\mathbf 0}}
\def\bGam{{\mathbf\Gamma}}
\def\bSg{{\mathbf\Sigma}}

\def\cR{{\cal R}}

\def\bra{\langle}
\def\ket{\rangle}
\def\up{\uparrow}
\def\down{\downarrow}

\def\alf{\alpha}
\def\eps{\epsilon}
\def\gam{\gamma}
\def\Gam{\Gamma}
\def\lam{\lambda}
\def\Lam{\Lambda}
\def\om{\omega}

\def\sg{\sigma}

\def\psib{\bar\psi}
\def\Psib{\bar\Psi}
\def\Phib{\bar\Phi}

\def\dag{\dagger}
\def\pdag{\phantom\dagger}

\usepackage[]{psfrag}

\psfrag{alpha}{$\alpha$}
\psfrag{gap}{$\Delta$}
\psfrag{lambda}{$\lambda$}
\psfrag{L}{$\Lambda$}
\psfrag{L_crit}{$\Lambda_c$}
\psfrag{alphgap}{$\alpha\,,\,\,\Delta$}
\psfrag{Lc_alph_gap}{$\Lambda_{c}\,,\,\,\alpha\,,\,\,\Delta$}

\psfrag{m_bosonic}{$m^{2}_{b},\,\,m^{2}_{\sigma}$}
\psfrag{Z_factors}{$Z_{\pi},\,\,Z_{\sigma}$}
\psfrag{g_factors}{$g_{\pi},\,\,g_{\sigma}$}

\begin{document}

\title{Renormalization group flow for fermionic superfluids at zero temperature}
\author{P. Strack}
\email{p.strack@fkf.mpg.de}
\affiliation{Max-Planck-Institute for Solid State Research,
 Heisenbergstr.\ 1, D-70569 Stuttgart, Germany} 
\affiliation{Institute for Theoretical Physics, University of Heidelberg, 
 Philosophenweg 16, D-69120 Heidelberg, Germany}
\author{R. Gersch}
\affiliation{Max-Planck-Institute for Solid State Research,
 Heisenbergstr.\ 1, D-70569 Stuttgart, Germany} 
\author{W. Metzner}
\affiliation{Max-Planck-Institute for Solid State Research,
 Heisenbergstr.\ 1, D-70569 Stuttgart, Germany} 
\date{\today}
\begin{abstract}
We present a comprehensive analysis of quantum fluctuation effects 
in the superfluid ground state of an attractively interacting Fermi 
system, employing the attractive Hubbard model as a prototype.
The superfluid order parameter, and fluctuations thereof, are implemented 
by a bosonic Hubbard-Stratonovich field, which splits into two components 
corresponding to longitudinal and transverse (Goldstone) fluctuations. 
Physical properties of the system are computed from a set of approximate 
flow equations obtained by truncating the exact functional renormalization 
group flow of the coupled fermion-boson action. 
The equations capture the influence of fluctuations on non-universal
quantities such as the fermionic gap, as well as the universal infrared 
asymptotics present in every fermionic superfluid.
We solve the flow equations numerically in two dimensions and compute 
the asymptotic behavior analytically in two and three dimensions. 
The fermionic gap $\Delta$ is reduced significantly compared to the 
mean-field gap, and the bosonic order parameter $\alpha$, which is 
equivalent to $\Delta$ in mean-field theory, is suppressed to values 
below $\Delta$ by fluctuations. 
The fermion-boson vertex is only slightly renormalized. 
In the infrared regime, transverse order parameter fluctuations 
associated with the Goldstone mode lead to a strong renormalization of 
longitudinal fluctuations: 
the longitudinal mass and the bosonic self-interaction vanish linearly 
as a function of the scale in two dimensions, and logarithmically in 
three dimensions, in agreement with the exact behavior of an interacting 
Bose gas.

\end{abstract}
\pacs{05.10.Cc, 67.25.D-, 71.10.Fd}

\maketitle

\section{Introduction}

Interacting Fermi systems exhibit very diverse behavior on 
different energy scales. Composite objects and collective phenomena
emerge at scales far below the bare energy scales of the microscopic
degrees of freedom.
This diversity of scales is a major obstacle to a numerical solution
of microscopic models, since the most interesting phenomena appear 
only at low temperatures and in systems with very large size.
It is also hard to deal with by conventional many-body methods, if one 
tries to treat all scales at once and within the same approximation,
such as a partial resummation of Feynman diagrams.

Therefore, it seems natural to integrate degrees of freedom (bosonic and/or 
fermionic fields) with different energy scales successively, descending
step by step from the highest scale present in the microscopic system.
This generates a one-parameter family of effective actions which 
interpolates smoothly between the bare action of the system, as given
by the microscopic Hamiltonian, and the final effective action from
which all physical properties can be extracted.
The vertex functions corresponding to the effective action at scale
$\Lam$ obey an exact hierarchy of differential flow equations, 
frequently referred to as ''exact'' or ''functional'' renormalization
group (fRG).\cite{berges_review02}
This hierarchy of flow equations is a transparent source of truncation 
schemes for a given physical problem.\cite{metzner05}

Most interacting Fermi systems undergo a phase transition associated
with spontaneous symmetry breaking at low temperature.
In a renormalization group flow,
the instability of the normal state is signalled by a divergence of
the effective two-particle interaction at a finite scale $\Lam_c \,$,
that is, before all degrees of freedom have been integrated out.
To continue the flow below the scale $\Lam_c$, the order parameter 
corresponding to the broken symmetry has to be implemented.
This can be accomplished in a purely fermionic theory or, alternatively, by 
introducing a bosonic order parameter field via a Hubbard-Stratonovich 
transformation.

In a fermionic description the order parameter can be implemented by
adding an infinitesimal external symmetry breaking field to the bare
Hamiltonian, which is then promoted to a finite order parameter below
the scale $\Lam_c \,$.\cite{salmhofer04} 
To capture first order transitions, one has to implement the order 
parameter via a finite counterterm, which is added to the bare 
Hamiltonian and then subtracted again in the course of the flow.
\cite{gersch06}
A relatively simple one-loop truncation \cite{katanin04} of
the exact fermionic flow equation hierarchy yields an exact
solution of mean-field models.\cite{salmhofer04,gersch05,gersch06}
The same truncation with momentum-dependent vertices provides a promising approximation to treat 
spontaneous symmetry breaking also beyond mean-field theory.
Recently it has been shown that this approximation yields quite 
accurate results for the fermionic gap in the ground state 
of the two-dimensional attractive Hubbard model for weak and moderate 
interactions.\cite{gersch08}

If the order parameter is implemented via a Hubbard-Stratonovich 
field, one has to deal with a coupled theory of bosons and fermions.
The evolution of the effective action of this theory is again given 
by an exact hierarchy of flow equations for the vertex functions.
\cite{baier04,schuetz05,schuetz06}
A truncation of this hierarchy has been applied a few years ago to 
the antiferromagnetic state of the two-dimensional repulsive Hubbard 
model.\cite{baier04} Important features of the quantum
antiferromagnet at low temperatures were captured by
the flow.
More recently, various aspects of superfluidity in attractively 
interacting Fermi systems have been studied in the fRG framework with
a Hubbard-Stratonovich field for the superfluid order parameter.
Approximate flow equations were discussed for the superfluid ground 
state,\cite{krippa07} for the Kosterlitz-Thouless transition in 
two-dimensional superfluids \cite{krahl07} and for the BCS-BEC crossover 
in three-dimensional cold atomic Fermi gases.\cite{diehl07}

In this work we analyze the fRG flow of the {\em attractive} Hubbard
model, using the Hubbard-Stratonovich route to symmetry breaking.
The attractive Hubbard model is a prototype of a Fermi system with
a superfluid low temperature phase.\cite{micnas90}
In particular, it is a popular model for the crossover from BCS-type 
superfluidity at weak coupling to Bose-Einstein condensation of 
strongly bound pairs at strong coupling.\cite{randeria95,keller01}
An experimental realization of the attractive Hubbard model is
conceivable by trapping cold fermionic atoms in an optical lattice
and tuning the interaction close to a Feshbach resonance.
\cite{hofstetter02,jaksch05,chin06}

We focus on the superfluid ground state.
The importance of quantum fluctuations in the superfluid ground
state has been emphasized recently in the context of the BCS-BEC
crossover.\cite{diener08}
Although the long-range order is not destroyed by fluctuations in
dimensions $d>1$, the order parameter correlations are nevertheless 
non-trivial in $d \leq 3$. 
The Goldstone mode leads to severe infrared divergences in perturbation 
theory. A detailed analysis of the infrared behavior of fermionic 
superfluids has appeared earlier in the mathematical literature,
\cite{feldman93} where the perturbative renormalizability of the 
singularities associated with the Goldstone mode was established 
rigorously.
To a large extent divergences of Feynman diagrams cancel due to 
Ward identities, while the remaining singularities require a 
renormalization group treatment.
Since the fermions are gapped at low energy scales, the infrared 
behavior of the collective, bosonic sector in fermionic superfluids 
is equivalent to the one of an interacting Bose gas, where
the Goldstone mode of the condensed state strongly affects the 
longitudinal correlations, leading to drastic deviations from
mean-field theory in dimensions $d \leq 3$.
\cite{nepomnyashchy92,pistolesi04}

The purpose of our work is to construct a relatively simple 
truncation of the exact fRG flow which is able to describe the
correct infrared asymptotic behavior, and which yields reasonable
estimates for the order parameter at least for weak and moderate 
interaction strength. From a numerical solution of the flow 
equations, which we perform in two dimensions, we obtain 
information on the importance of Goldstone modes
and other fluctuation effects.

In Sec.~II we introduce the bare fermion-boson action obtained from
the attractive Hubbard model by a Hubbard-Stratonovich transformation.
Neglecting bosonic fluctuations, one recovers the standard mean-field
theory for fermionic superfluids, as recapitulated in Sec.~III.
By truncating the exact fRG hierarchy, we derive approximate flow 
equations involving fermionic and bosonic fluctuations in Sec.~IV. 
At the end of that section we reconsider mean-field theory from a flow 
equation perspective. 
Sec.~V is dedicated to a discussion of results obtained by solving
the flow equations. 
We discuss the asymptotic behavior in the infrared limit in two and 
three dimensions and then present numerical results for the flow
in two dimensions, where fluctuation effects are most pronounced.
Finally, we summarize our results 
in Sec.~VI.

\section{Bare action}

As a prototype model for the formation of a superfluid ground state in an
interacting Fermi system we consider the attractive Hubbard model
\begin{equation}
 H = \sum_{\bi,\bj} \sum_{\sg} t_{\bi\bj} \,
 c^{\dag}_{\bi\sg} c^{\pdag}_{\bj\sg} +
 U \sum_{\bi} n_{i\up} n_{i\down} \; ,
\end{equation}
where $c^{\dag}_{\bi\sg}$ and $c^{\pdag}_{\bi\sg}$ are creation and 
annihilation operators for spin-$\frac{1}{2}$ fermions with spin orientation 
$\sg$ on a lattice site $\bi$.
For the hopping matrix we employ $t_{\bi\bj} = - t$ if $\bi$ and $\bj$
are nearest neighbors on the lattice, and $t_{\bi\bj} = 0$ otherwise.
On a $d$-dimensional simple cubic lattice, this leads to a dispersion 
relation $\eps_{\bk} = -2t \sum_{i=1}^d \cos k_i$.
For the attractive Hubbard model the coupling constant $U$ is
negative.

The attractive Hubbard model has a superfluid ground state for any 
particle density $n$ in $d \geq 2$ dimensions,\cite{micnas90} provided the
lattice is not completely filled ($n=2$) or empty ($n=0$).
At half filling ($n=1$) the usual U(1) global gauge symmetry becomes a 
subgroup of a larger SO(3) symmetry group, and the order parameter for
superfluidity mixes with charge density order.\cite{micnas90}

Our analysis is based on a functional integral representation of the
effective action, that is, the generating functional of one-particle
irreducible correlation functions.
For the Hubbard model, the starting point is a functional integral
over fermionic fields $\psi$ and $\psib$ with the {\em bare} action
\begin{eqnarray}
  \Gam_0[\psi,\psib]
  & = & - \int_{k\sg}
  \psib_{k\sg} (ik_{0}-\xi_{\bk}) \, \psi_{k\sg} \nonumber \\
  && + \int_{k,k',q} U \, 
  \psib_{-k+\frac{q}{2}\down} \psib_{k+\frac{q}{2} \up}
  \psi_{k'+\frac{q}{2}\up} \psi_{-k'+\frac{q}{2}\down} \; ,
  \label{eq:bare_action}
\end{eqnarray}
where $\xi_{\bk} = \eps_{\bk} - \mu$ is the single-particle energy
relative to the chemical potential.
The variables $k = (k_0,\bk)$ and $q = (q_0,\bq)$
collect Matsubara energies and momenta. 
We use the short-hand notation
$\int_k = \int_{k_0} \int_{\bk} = 
 \int_{-\infty}^{\infty} \frac{d k_{0}}{2\pi}
 \int_{-\pi}^{\pi} \frac{d^d \bk}{(2\pi)^d} \,$
for momentum and energy integrals, 
and $\int_{k\sg}$ includes also a spin sum. 
We consider only {\em ground state} properties, such that
the energy variables are continuous.

The attractive interaction drives spin-singlet pairing with s-wave
symmetry and a spontaneous breaking of the global U(1) 
gauge symmetry.
Therefore, we decouple the Hubbard interaction in the s-wave 
spin-singlet pairing channel by introducing a complex bosonic 
Hubbard-Stratonovich field $\phi_q$ conjugate to the bilinear 
composite of fermionic fields \cite{popov87}
\begin{equation}
 \tilde\phi_{q} = U \int_k \psi_{k+\frac{q}{2}\up}
 \psi_{-k+\frac{q}{2}\down} \; .
\end{equation}
This yields a functional integral over $\psi$, $\psib$ and $\phi$
with the new bare action
\begin{eqnarray}
 \Gam_0[\psi,\psib,\phi]
  &=& - \int_{k\sg} \psib_{k\sg} (ik_{0} - \xi_{\bk}) \, \psi_{k\sg}
  - \int_q \phi^*_q \frac{1}{U} \phi_q \nonumber \\
  && + \int_{k,q} \left( 
  \psib_{-k+\frac{q}{2} \down} \psib_{k+\frac{q}{2} \up} \,
  \phi_{q} +
  \psi_{k+\frac{q}{2} \up} \psi_{-k+\frac{q}{2}\down} \,
  \phi^*_q \right) . \hskip 5mm
  \label{eq:finalmodel}
\end{eqnarray}
where $\phi^*$ is the complex conjugate of $\phi$, while $\psi$
and $\psib$ are algebraically independent Grassmann variables.

Our aim is to compute fermionic and bosonic correlation functions
with a focus on a correct description of the low-energy (infrared)
behavior. A central object in our analysis is the {\em effective
action}\/ $\Gam[\psi,\psib,\phi]$, which can be obtained by 
functional integration of the bare action in the 
presence of source fields and a subsequent 
Legendre transform with respect to these fields.\cite{negele87}
Functional derivatives of $\Gam[\psi,\psib,\phi]$ with 
respect to $\psi$, $\psib$, $\phi$ (and $\phi^*$) yield the 
one-particle irreducible vertex functions.

\section{Mean-field theory}

As a warm-up for the renormalization group treatment it is instructive
to recapitulate the mean-field theory for the superfluid phase
in the functional integral formalism.\cite{popov87}
In mean-field approximation bosonic fluctuations are neglected, 
that is, the bosonic field $\phi$ is fixed instead of being integrated 
over all possible configurations. 
The fermion fields can then be integrated exactly. 
The (fixed) bosonic field is determined by minimizing the effective 
action as a functional of $\phi$.  
For a homogeneous system, the minimizing $\phi_q$ can be be non-zero
only for $q = 0$. We denote the minimum by $\alf$.
Substituting $\phi_0 \to \alf + \phi_0$ yields
\begin{eqnarray}
 \Gam_0[\psi,\psib,\alf+\phi] \! &=& 
 - \int_{k\sg} \psib_{k\sg} (ik_{0} - \xi_{\bk}) \, \psi_{k\sg}
 - \alf^* \frac{1}{U} \alf 
 \nonumber \\
 && + \int_k \left( \psib_{-k\down} \psib_{k\up} \, \alf
  + \psi_{k\up} \psi_{-k\down} \, \alf^* \right)
 \nonumber \\
 && - \frac{1}{U} 
 \left( \alf^* \phi_0 + \alf \, \phi_0^* \right)
 \nonumber \\
 && + \int_{k,q} \left( 
 \psib_{-k+\frac{q}{2} \down} \psib_{k+\frac{q}{2} \up} \,
 \phi_{q} +
 \psi_{k+\frac{q}{2} \up} \psi_{-k+\frac{q}{2}\down} \,
 \phi^*_q \right)
 \nonumber \\
 && - \int_q \phi_q^* \frac{1}{U} \phi_q \; .
\label{eq:mft_Gam_0}
\end{eqnarray}
A necessary condition for a minimum of the effective action is
that its first derivative with respect to $\phi$ (or $\phi^*$),
that is, the bosonic 1-point function $\Gam_b^{(1)}(q)$, vanishes.
In other words, terms linear in $\phi$ (or $\phi^*$) have to vanish
in the effective action.
For $q \neq 0$, $\Gam_b^{(1)}(q)$ vanishes for any choice of $\alf$
in a homogeneous system.
For $q = 0$ and in mean-field approximation, the 1-point function is
given by
\begin{equation}
 \Gam_b^{(1)}(0) = - \frac{1}{U} \alf + 
 \int_k \bra \psi_{k\up} \psi_{-k\down} \ket
\label{eq:orderparam_mft}
\end{equation}
where $\bra \dots \ket$ denotes expectation values.
The first term on the right hand side corresponds to the contribution
$- \frac{1}{U} \alf \phi_0^*$ to $\Gam_0$ in the third line of Eq.~(\ref{eq:mft_Gam_0}), 
while the second term is generated by contracting the fermions in the
contribution proportional to $\phi_q^*$ in the forth line of Eq.~(\ref{eq:mft_Gam_0}).
In the absence of bosonic fluctuations there is no other contribution
to $\Gam_b^{(1)}$.
From the condition $\Gam_b^{(1)}(0) = 0$ one obtains
\begin{equation}
 \alf = U \int_k \bra \psi_{k\up} \psi_{-k\down} \ket \; ,
\end{equation}
which relates $\alf$ to a fermionic expectation value.

We now turn to the fermionic 2-point functions. 
The normal fermionic propagator
$G_{f\sg}(k) = - \bra \psi_{k\sg} \psib_{k\sg} \ket$ 
and the anomalous propagators
$F_f(k) = - \bra \psi_{k\up} \psi_{-k\down} \ket$ and
$\bar F_f(k) = - \bra \psib_{-k\down} \psib_{k\up} \ket$
can be conveniently collected in a Nambu matrix propagator
\begin{equation}
 \bG_f(k) = 
 \left( \begin{array}{cc}
 G_{f\up}(k) & F_f(k) \\[1mm]
 \bar F_f(k) & - G_{f\down}(-k)
 \end{array} \right) \; .
\end{equation}
The anomalous propagators satisfy the relations
$\bar F_f(k) = F_f^*(k)$ and $F_f(-k) = F_f(k)$.
In (our) case of spin rotation invariance the normal propagator
does not depend on $\sg$, and therefore, 
$G_{f\up}(k) = G_{f\down}(k) = G_f(k)$.

In mean-field theory, the fermionic 2-point vertex function
$\bGam_f^{(2)} = - \bG_f^{-1}$
can be read off directly from the bare action in the form Eq.~(\ref{eq:mft_Gam_0}):
\begin{equation}
 \bGam_f^{(2)}(k) = - \left( \begin{array}{cc}
 ik_0 - \xi_{\bk} & \alf \\
 \alf^* & ik_0 + \xi_{-\bk}
 \end{array} \right) \; .
\end{equation}
The off-diagonal elements are due to the terms in the second line 
of Eq.~(\ref{eq:mft_Gam_0}). Tadpole contributions which are generated from the terms
in the lines 3-5 of Eq.~(\ref{eq:mft_Gam_0}) cancel exactly by virtue of the condition 
$\Gam_b^{(1)} = 0$. In the absence of bosonic fluctuations there are
no other contributions to $\bGam_f^{(2)}$.
Inverting $\bGam_f^{(2)}$ and using $\xi_{-\bk} = \xi_{\bk}$ from 
reflection symmetry yields
\begin{eqnarray}
 G_f(k) &=& 
 \frac{-ik_0 - \xi_{\bk}}{k_0^2 + E_{\bk}^2}
 \label{eq:Gf}
 \\
 F_f(k) &=& 
 \frac{\Delta}{k_0^2 + E_{\bk}^2} \; ,
\label{eq:Ff}
\end{eqnarray}
where $E_{\bk} = (\xi_{\bk}^2 + |\Delta|^2 )^{1/2}$ and 
$\Delta = \alf$.
We observe that in mean-field theory the bosonic order parameter $\alf$ 
is equivalent to the gap $\Delta$ in the fermionic excitation spectrum.
Eq.~(\ref{eq:orderparam_mft}) corresponds to the BCS gap equation
\begin{equation}
 \Delta = - U \int_k F_f(k) \; .
\label{eq:bcs_gap_eqn}
\end{equation}

We finally compute the bosonic 2-point functions in mean-field theory.
The bosonic propagators $G_b(q) = - \bra \phi_q \phi^*_q \ket$ and 
$F_b(q) = - \bra \phi_q \phi_{-q} \ket$ = 
$- \bra \phi^*_{-q} \phi^*_q \ket^*$ form the matrix propagator
\begin{equation}
 \bG_b(q) = 
 \left( \begin{array}{cc}
 G_b(q) & F_b(q) \\[1mm]
 F_b^*(q) & G_b(-q)
 \end{array} \right) \; .
\end{equation}
Note that $F_b(-q) = F_b(q)$.
The bosonic 2-point function $\bGam_b^{(2)}$ is equal to
$-\bG_b^{-1}$. 
We define a bosonic self-energy $\bSg_b$ via the Dyson equation
$(\bG_b)^{-1} = (\bG_{b0})^{-1} - \bSg_b$, where the bare propagator
corresponding to the bare action $\Gam_0$ is given by
\begin{equation}
 \bG_{b0}(q) = 
 \left( \begin{array}{cc}
 U & 0 \\
 0 & U
 \end{array} \right) \; .
\end{equation}

In mean-field theory, only fermionic bubble diagrams contribute to
the bosonic self-energy:
\begin{equation}
 \bSg_b(q) = \left( \begin{array}{cc}
 K(q) & L(q) \\
 L^*(q) & K(-q)
 \end{array} \right) \; ,
\end{equation}
where
\begin{eqnarray}
 K(q) &=& - \int_k G_f(k+q) \, G_f(-k) \\[2mm]
 L(q) &=& {\phantom -} \int_k F_f(k+q) \, F_f(-k)\;.
\end{eqnarray}
In the absence of bosonic fluctuations, there are no other contributions 
to $\bSg_b$. Tadpole diagrams cancel due to $\Gam_b^{(1)} = 0$.
Note that $K(-q) = K^*(q)$ while $L(-q) = L(q)$.
Inverting the matrix $\bG_{b0} - \bSg_b$ one obtains the bosonic 
propagator in mean-field approximation
\begin{equation}
 \bG_b(q) = \frac{1}{d(q)} \left( \begin{array}{cc}
 U^{-1} - K(-q) & L(q) \\
 L^*(q) & U^{-1} - K(q)
 \end{array} \right) \; ,
\end{equation}
with the determinant
$d(q) = |U^{-1} - K(q)|^2 - |L(q)|^2$.

Using the explicit expressions (\ref{eq:Gf}) and (\ref{eq:Ff}) 
for $G_f$ and $F_f$, respectively, one can see that
\begin{equation}
 U^{-1} - K(0) + |L(0)| = U^{-1} + \frac{1}{\Delta} \int_k F_f(k) \; ,
\end{equation}
which vanishes if $\Delta$ is non-zero and satisfies the gap equation.
Hence $d(q)$ has a zero and $\bG_b(q)$ a pole in $q=0$.
This pole corresponds to the Goldstone mode associated with the
sponaneous breaking of the U(1) symmetry of the model.
For small finite $\bq$ and $q_0$, the leading $q$-dependences of
$|U^{-1} - K(q)|$ and $L(q)$ are of order $|\bq|^2$ and $q_0^2$.
Hence the divergence of $\bG_b(q)$ for $q \to 0$ is quadratic in
$\bq$ and $q_0$. Continuing $q_0$ to real frequencies one obtains
a propagating mode with a linear dispersion relation. As expected, the second pole 
of $d(q)$ is gapped and features a quadratic momentum dispersion. 

By appropriately tailoring the fRG-trunction in Sec. \ref{subsubsec:trunc_ssb} to the 
results of this mean-field calculation in the superfluid phase, 
we incorporate the effects of transversal (Goldstone) fluctuations as well as 
longitudinal (radial) fluctuations into our computation.

\section{Flow equations}

In this section we derive approximate flow equations by truncating 
the exact flow equation\cite{wetterich93} for the effective action.

\subsection{Exact flow equation}

To simplify our notation, we use fermionic Nambu fields
\begin{equation}
 \Psi_k = \left( \begin{array}{c}
 \psi_{k\up} \\ \psib_{-k\down}
 \end{array} \right) \; , \quad
 \Psib_k = \left( \psib_{k\up}, \psi_{-k\down} \right)
\end{equation}
and bosonic Nambu fields
\begin{equation}
 \Phi_q = \left( \begin{array}{c}
 \phi_q \\ \phi^*_{-q}
 \end{array} \right) \; , \quad
 \Phib_q = 
 \left( \phi^*_q, \phi_{-q} \right) \; .
\end{equation}
The fermionic and bosonic matrix propagators are then given by
$\bG_f(k) = - \bra \Psi_k \Psib_k \ket$ and
$\bG_b(q) = - \bra \Phi_q \Phib_q \ket \,$, respectively.
To write down the exact flow equations it is convenient to combine
fermionic and bosonic fields in a superfield $\mathcal{S}$, where fermions
and bosons are distinguished by a statistics index $s= b,f$, that is, 
$\mathcal{S}_b = \Phi$ and $\mathcal{S}_f = \Psi$. The superpropagator 
$\bG(q) = - \bra \mathcal{S}_q \bar{\mathcal{S}}_q \ket$ is diagonal in the statistics 
index.

The flow equation describes the evolution of the effective action as
a function of a flow parameter $\Lam$, usually related to a cutoff.
In this work we use sharp frequency cutoffs which exclude bosonic
fields with $|\mbox{frequency}| < \Lam_b$ and fermionic fields with 
$|\mbox{frequency}| < \Lam_f$ from the functional integral. 
Thereby both fermionic and bosonic infrared divergences are 
regularized.
Both cutoffs are monotonic functions of the flow parameter,
$\Lam_b(\Lam)$ and $\Lam_f(\Lam)$, which vanish for $\Lam \to 0$ and
tend to infinity for $\Lam \to \infty$.
The cutoff can be implemented by adding the regulator term
\begin{equation}
 \cR^{\Lam} = 
 \frac{1}{2} \int_q \Phib_q \, \bR_b^{\Lam}(q) \, \Phi_q +
 \int_k \Psib_k \, \bR_f^{\Lam}(k) \, \Psi_k
\end{equation}
to the bare action, where
\begin{equation}
 \bR_s^{\Lam}(k) = [\bG_{s0}(k)]^{-1} -
 [\chi_s^{\Lam}(k_0) \, \bG_{s0}(k)]^{-1}
\end{equation}
for $s = b,f$, and $\chi_s^{\Lam}(k_0) = \Theta(|k_0| - \Lam_s)$. 
This term replaces the bare propagators $\bG_{s0}$ by 
$\bG_{s0}^{\Lam} = \chi_s^{\Lam} \bG_{s0}$.

Integrating $e^{-\Gam_0 - \cR^{\Lam}}$ in the presence of source 
fields coupling linearly to $\mathcal{S}$ and $\bar{\mathcal{S}}$ yields the 
cutoff-dependent generating functional for connected Green 
functions
\begin{equation}
 G^{\Lam}[\mathcal{S}',\bar{\mathcal{S}}'] = - \log \int D\mathcal{S} D\bar{\mathcal{S}} \, 
 e^{-\Gam_0[\mathcal{S},\bar{\mathcal{S}}] - \cR^{\Lam}[\mathcal{S},\bar{\mathcal{S}}] + 
 (\mathcal{S}',\bar{\mathcal{S}}) + (\mathcal{S},\bar{\mathcal{S}}')} \; ,
\end{equation}
where the bracket $(.,.)$ is a shorthand notation for the inner product for superfields.
The cutoff-dependent effective action $\Gam^{\Lam}$ is defined as
\begin{equation}
 \Gam^{\Lam}[\mathcal{S},\bar{\mathcal{S}}] = 
 {\cal L} G^{\Lam}[\mathcal{S},\bar{\mathcal{S}}] - \cR^{\Lam}[\mathcal{S},\bar{\mathcal{S}}] \; ,
\end{equation}
where ${\cal L} G^{\Lam}$ is the Legendre transform of $G^{\Lam}$.
As a function of decreasing cutoff,
$\Gam^{\Lam}$ interpolates smoothly between the bare 
action $\Gam_0$ for $\Lam = \infty$ and the full effective 
action $\Gam$ recovered in the limit $\Lam \to 0$.

The evolution of the effective action follows the exact flow
equation 
\begin{equation}
 \frac{d}{d\Lam} \Gam^{\Lam}[\mathcal{S},\bar{\mathcal{S}}] = 
 {\rm Str} \, \frac{\dot{\bR}^{\Lam}}
 {\bGam^{(2) \, \Lam}[\mathcal{S},\bar{\mathcal{S}}] + \bR^{\Lam}} \; ,
\label{eq:exact_flow_eqn}
\end{equation}
where $\dot{\bR}^{\Lam} = \partial_{\Lam} \bR^{\Lam}$, and
\begin{equation}
 \bGam^{(2) \,\Lam}[\mathcal{S},\bar{\mathcal{S}}] =
 \frac{\partial^2 \Gam^{\Lam}[\mathcal{S},\bar{\mathcal{S}}]}
 {\partial\mathcal{S} \partial\bar{\mathcal{S}}} \; .
\end{equation}
The supertrace $\rm Str$ traces over all indices with a plus 
sign for bosons and a minus sign for fermions.
Note that the definitions of $\Gam^{\Lam}$ vary slightly in the
literature. In particular, $\Gam^{\Lam}$ is frequently defined
as the Legendre transform of $G^{\Lam}$ without subtracting
the regulator term $\cR^{\Lam}$, which leads to a simple 
additional term in the flow equation.

To expand the functional flow equation (\ref{eq:exact_flow_eqn}) in powers of the
fields, we write the Hessian of $\Gam^{\Lam}$ as
\begin{equation}
 \bGam^{(2) \,\Lam}[\mathcal{S},\bar{\mathcal{S}}] = 
 - (\bG^{\Lam})^{-1} + \tilde\bGam^{(2) \,\Lam}[\mathcal{S},\bar{\mathcal{S}}] \; ,
\end{equation}
where $\tilde\bGam^{(2) \,\Lam}[\mathcal{S},\bar{\mathcal{S}}]$ contains only terms
which are at least quadratic in the fields.
Defining $\bG_R^{\Lam} = [(\bG^{\Lam})^{-1} - \bR^{\Lam}]^{-1}$
and expanding in powers of $\tilde\bGam^{(2) \,\Lam}$ yields
\begin{eqnarray}
 \frac{d}{d\Lam} \Gam^{\Lam} &=& 
 - {\rm Str} (\dot{\bR}^{\Lam} \bG_R^{\Lam}) \nonumber \\
 &-& {\rm Str} \left[
 {\bG'_R}^{\!\Lam} \big( \tilde\bGam^{(2) \,\Lam} + 
 \tilde\bGam^{(2) \,\Lam} \bG_R^{\Lam} \tilde\bGam^{(2) \,\Lam} + \dots
 \big) \right] \; , \quad
\label{eq:flow_expansion}
\end{eqnarray}
where
\begin{equation}
 {\bG'_R}^{\!\Lam} = 
 \bG_R^{\Lam} \dot{\bR}^{\Lam} \bG_R^{\Lam} \; .
\end{equation}
The socalled single-scale propagator ${\bG'_R}^{\!\Lam}$ has 
support only for frequencies at the cutoffs, that is, for 
$|k_0| = \Lam_s$. 
Expanding both sides of Eq.~(\ref{eq:flow_expansion}) in powers of the fields and
comparing coefficients yields a hierarchy of flow equations 
for the vertex functions.

To describe spontaneous symmetry breaking we expand around a
generally non-zero value of the $q=0$ component of the bosonic 
field, that is, we expand
$\Gam^{\Lam}[\psi,\psib,\alf^{\Lam}+\phi]$ 
in powers of $\psi,\psib,\phi,\phi^*$, where $\alf^{\Lam}$ 
may be non-zero.
The expansion point $\alf^{\Lam}$ will be chosen such that the 
bosonic 1-point function $\Gam_b^{(1)\Lam}$ vanishes for all 
$\Lam$. In this way tadpole contributions are avoided.
The $\Lam$-dependence of the expansion point $\alf^{\Lam}$
leads to terms proportional to 
$\dot{\alf}^{\Lam} = \partial_{\Lam} \alf^{\Lam}$ stemming
from the total $\Lam$-derivative on the left hand side of 
the flow equation.

In the following we will frequently suppress the superscript 
$\Lam$ attached to cutoff-dependent quantities.

\subsection{Truncation}

The exact effective action contains an infinite number of
terms of arbitrary order in fermionic and/or bosonic fields.
We now describe which terms are kept and how they are 
parametrized.
We keep all terms which are crucial for a qualitatively
correct description of the low-energy behavior of the
system.
We distinguish the symmetric regime, where $\alf = 0$,
from the symmetry broken regime, where $\alf \neq 0$.
The former applies to large $\Lam$, the latter to small 
$\Lam$.

\subsubsection{Symmetric regime}
\label{subsubsec:trunc_sym}
The bare action Eq.~(\ref{eq:finalmodel}) contains quadratic 
terms for fermions and bosons, and an interaction term where 
bosons couple linearly to a fermion bilinear. In the effective 
action we keep these terms with generalized cutoff-dependent
parameters and add a bosonic self-interaction of order
$|\phi|^4$. The latter is generated by the flow and 
becomes crucial when the quadratic part of the bosonic
potential changes sign.
Other interactions generated by the flow 
are neglected.

For our choice of parameters (relatively small $U$), the 
fermionic propagator receives only Fermi liquid 
renormalizations, leading to a slightly reduced quasi particle
weight and a weakly renormalized dispersion relation. 
We neglect these quantitatively small effects and leave 
the quadratic fermionic term in the action unrenormalized,
that is,
\begin{equation}
 \Gam_{\psib\psi} = - \int_{k\sg} 
 \psib_{k\sg} (ik_0 - \xi_{\bk}) \, 
 \psi_{k\sg} \; ,
\label{eq:Gam_psi_psi}
\end{equation}
corresponding to an unrenormalized fermionic propagator
\begin{equation}
 G_{f}(k) = G_{f0}(k) =
 \frac{1}{ik_{0}-\xi_{\mathbf{k}}} \; . 
\end{equation}
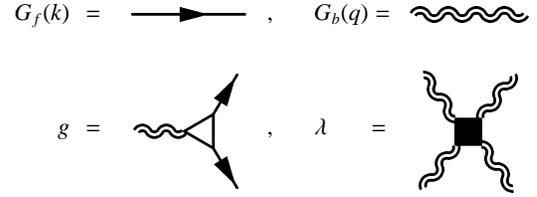
\begin{figure}
\begin{fmffile}{legend_sym_27}
\begin{eqnarray}
G_{f}(k)&=&
\parbox{20mm}{\unitlength=1mm\fmfframe(1,1)(1,1){
\begin{fmfgraph*}(15,1)\fmfpen{thin} 
\fmfleft{l1}
 \fmfright{r1}
  \fmf{fermion}{l1,r1}
 \end{fmfgraph*}
}}\nonumber,\hspace{5mm}
G_{b}(q)=
\parbox{20mm}{\unitlength=1mm\fmfframe(1,1)(1,1){
\begin{fmfgraph*}(15,1)\fmfpen{thin} 
\fmfleft{l1}
 \fmfright{r1}
  \fmf{dbl_wiggly}{l1,r1}
 \end{fmfgraph*}
}}\nonumber\\[3mm]
g&=&
\parbox{20mm}{\unitlength=1mm\fmfframe(2,2)(1,1){
\begin{fmfgraph*}(15,15)\fmfpen{thin}
\fmfleft{l1}
\fmfrightn{r}{2}
\fmf{dbl_wiggly}{l1,G1}
\fmfpolyn{empty,tension=0.8}{G}{3}
\fmf{fermion}{G2,r1}
\fmf{fermion}{G3,r2}
 \end{fmfgraph*}
}},\nonumber\hspace{5mm}
\lambda\hspace{5mm}=
\parbox{20mm}{\unitlength=1mm\fmfframe(2,2)(1,1){
\begin{fmfgraph*}(15,15)\fmfpen{thin}
\fmfleftn{l}{2}
\fmfrightn{r}{2}
\fmf{dbl_wiggly}{l2,G1}
\fmf{dbl_wiggly}{l1,G2}
\fmfpolyn{full,tension=1.5}{G}{4}
\fmf{dbl_wiggly}{r1,G3}
\fmf{dbl_wiggly}{r2,G4}
 \end{fmfgraph*}
}}\nonumber\\[-7mm]\nonumber
\end{eqnarray}
\end{fmffile}
\caption{Diagrammatic constituents of our truncation in the symmetric regime as 
described in Sec. \ref{subsubsec:trunc_sym}.}
\end{figure}
In the bare action the term quadratic in bosons contains
only a mass term. In the effective action this mass 
decreases with decreasing cutoff until it vanishes at
a critical scale $\Lam_c$, which marks the transition to
the symmetry-broken regime. As the mass decreases, the
momentum and frequency dependence of the bosonic 2-point
function becomes important. The latter is generated in
particular by fermionic fluctuations. For small $\bq$, 
the leading $\bq$-dependence is of order $|\bq|^2$.
The leading frequency-dependent contribution to the
real part of the bosonic 2-point function is of order
$q_0^2$. The frequency-dependence of the imaginary part
is generally of order $q_0$, but the prefactor is very 
small, which is related to the fact that it vanishes
completely in case of particle-hole symmetry. Furthermore
this small imaginary part does not have any qualitative
impact on the quantities we compute in the following.
We therefore neglect this term and make the ansatz
\begin{equation}
 \Gam_{\phi^*\phi} = \frac{1}{2} \int_q \phi_q^*
 ( m_b^2 + Z_b q_0^2 + A_b \om_{\bq}^2 ) \, \phi_q \; ,
\end{equation}
where 
$\om_{\bq}^2 = 2 \sum_{i=1}^d \left(1-\cos q_i\right)$ is fixed,
while $m_{b}^{2}$, $Z_b$, and $A_b$ are cutoff-dependent numbers.
The function $\om_{\bq}^2$ has been chosen such that the
quadratic momentum dependence for small $\bq$ is 
continued to a periodic function defined on the entire
Brillouin zone.
The initial conditions for the parameters in the bosonic
2-point function,
\begin{eqnarray}
G_{b}(q)=-\frac{2}{Z_{b}q_{0}^{2}+A_{b}\omega^{2}_{\mathbf{q}}+m_{b}^{2}}\,\,, 
\end{eqnarray}
can be read off from the bare action as 
$m^{2}_b = |2/U|$ and $Z_b = A_b = 0$.

The interaction between fermions and bosons remains regular
and finite near $\Lam_c$. It can therefore be parametrized
as
\begin{equation}
 \Gam_{\psi^2\phi^*} = 
  g \int_{k,q} \left( 
  \psib_{-k+\frac{q}{2} \down} \psib_{k+\frac{q}{2} \up} \,
  \phi_{q} +
  \psi_{k+\frac{q}{2} \up} \psi_{-k+\frac{q}{2}\down} \,
  \phi^*_q \right) \; ,
\label{eq:g_sym}
\end{equation} 
where the coupling constant $g$ depends on the cutoff, but
not on momentum and frequency. The initial condition for 
$g$ is $g = 1$.

The flow generates a bosonic self-interaction which plays a
crucial role near and in the symmetry-broken regime, that is, 
when the bosonic mass term becomes small and finally changes 
sign. 
The most relevant term is a local $|\phi|^4$-interaction
\begin{equation}
 \Gam_{|\phi|^4} = \frac{\lam}{8} \int_{q,q',p}
 \phi^*_{q+p} \phi^*_{q'-p} \phi_{q'} \phi_q \; ,
\end{equation}
where the coupling constant $\lam$ depends on the cutoff 
but not on momentum and frequency.
The initial condition for $\lam$ is $\lam = 0$.

The propagators and interaction vertices are represented 
diagrammatically in Fig.~1.
\subsubsection{Symmetry broken regime}
\label{subsubsec:trunc_ssb}

For $\Lam < \Lam_c$ the effective action develops a 
minimum at $\phi_{q=0} = \alf \neq 0$. Due to the U(1)
symmetry associated with charge conservation the minimum
is degenerate with respect to the phase of $\alf$. 
In the following we choose $\alf$ real and positive.
We decompose the fluctuations of the bosonic field around
$\alf$ in a longitudinal (real) and a transverse 
(imaginary) part, $\sg$ and $\pi$, respectively:
\begin{eqnarray}
 \phi_q &=& \sg_q + i\pi_q \nonumber \\
 \phi_q^* &=& \sg_{-q} - i\pi_{-q} \; ,
\end{eqnarray}
where $\pi_{-q} = \pi_q^*$ and $\sg_{-q} = \sg_q^*$.

The bosonic part of the effective action consists of a 
local potential and momentum and frequency dependent
contributions to the 2-point functions.
A U(1)-symmetric local potential of order $|\phi|^4$ with
a minimum in $\alf$ has the form
\begin{equation}
 U^{\rm loc}[\phi] =
 \frac{\lam}{8} \int \left( |\phi|^2 - |\alf|^2 \right)^2 \; ,
\label{eq:U_local}
\end{equation}
where $\phi$ is the original (unshifted) bosonic field.
Substituting $\phi \to \alf + \sg + i\pi$ and expanding
in $\sg$ and $\pi$ yields a mass term for the $\sg$-field
and various interaction terms (see below).
No mass term for the $\pi$-field appears, as expected, 
since the $\pi$-field is a Goldstone mode.\cite{wetterich91}

The leading momentum and frequency dependence of the bosonic
2-point function is quadratic in $\bq$ and $q_0$, both for the
$\sg$- and $\pi$-component.
Hence we make the following ansatz for the quadratic bosonic 
contributions to the effective action
\begin{equation}
 \Gam_{\sg\sg} = \frac{1}{2} \int_q \sg_{-q}
 ( m_{\sg}^2 + Z_{\sg} q_0^2 + A_{\sg} \om_{\bq}^2 ) \, 
 \sg_q
\end{equation}
and
\begin{equation}
 \Gam_{\pi\pi} = \frac{1}{2} \int_q \pi_{-q}
 ( Z_{\pi} q_0^2 + A_{\pi} \om_{\bq}^2 ) \, \pi_q \; .
\end{equation}
where $m_{\sg}$, $Z_{\sg}$, $A_{\sg}$, $Z_{\pi}$, and 
$A_{\pi}$ are cutoff dependent real numbers. 
The propagators for the $\sg$ and $\pi$ fields thus have
the form
\begin{equation}
 G_{\sg}(q) = 
 - \frac{1}{m_{\sg}^2 + Z_{\sg} q_0^2 + A_{\sg} \om_{\bq}^2}
\end{equation}
and
\begin{equation}
 G_{\pi}(q) = 
 - \frac{1}{Z_{\pi} q_0^2 + A_{\pi} \om_{\bq}^2} \; ,
\end{equation}
respectively.
The longitudinal mass is determined by the $|\phi|^4$
coupling $\lam$ and the minimum $\alf$ as
\begin{equation}
 m_{\sg}^2 = \lam \, |\alf|^2 \; .
\label{eq:mSig_condition}
\end{equation}
The small imaginary contribution to $\Gam_{\phi^*\phi}$ generated 
by fermionic fluctuations in the absence of particle-hole symmetry,
mentioned already above, gives rise to an off-diagonal quadratic
term $\Gam_{\sg\pi}$ with a contribution linear in $q_0$. Since this
term would complicate the subsequent analysis without having any
significant effect, we will neglect it.

The bosonic interaction terms obtained by expanding
$U^{\rm loc}[\phi]$ are
\begin{eqnarray}
 \Gam_{\sg^4} &=& \gam_{\sg^4} \int_{q,q',p}
 \sg_{-q-p} \sg_{-q'+p} \sg_{q'} \sg_q \; , \\
 \Gam_{\pi^4} &=& \gam_{\pi^4} \int_{q,q',p}
 \pi_{-q-p} \pi_{-q'+p} \pi_{q'} \pi_q \; , \\
 \Gam_{\sg^2\pi^2} &=& \gam_{\sg^2\pi^2} \int_{q,q',p}
 \sg_{-q-p} \sg_{-q'+p} \pi_{q'} \pi_q \; , \\
 \Gam_{\sg^3} &=& \gam_{\sg^3} \int_{q,q'}
 \sg_{-q-q'} \sg_{q'} \sg_q \; , \\
 \Gam_{\sg\pi^2} &=& \gam_{\sg\pi^2} \int_{q,q'}
 \sg_{-q-q'} \pi_{q'} \pi_q \; ,
\end{eqnarray}
with $\gam_{\sg^4} = \gam_{\pi^4} = \lam/8$,
$\gam_{\sg^2\pi^2} = \lam/4$, and
$\gam_{\sg^3} = \gam_{\sg\pi^2} = \lam\alf/2$.
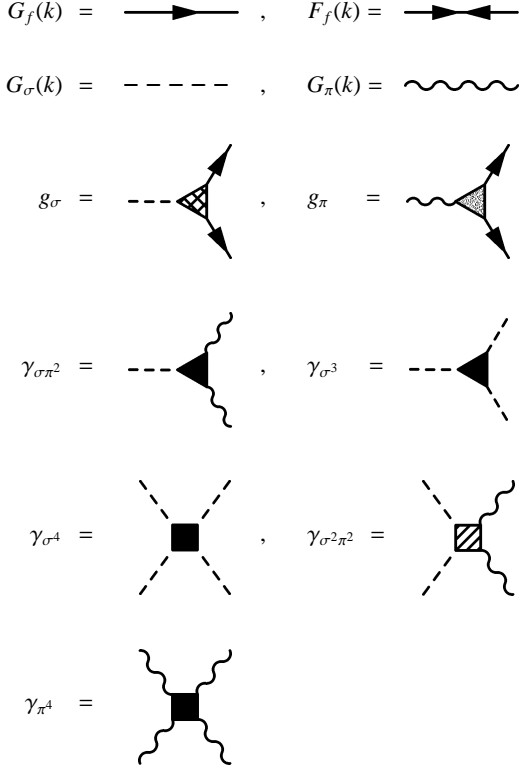
\begin{figure}
\begin{fmffile}{legend_ssb_15}
\begin{eqnarray}
G_{f}(k)&=&
\parbox{20mm}{\unitlength=1mm\fmfframe(1,1)(1,1){
\begin{fmfgraph*}(15,1)\fmfpen{thin} 
\fmfleft{l1}
 \fmfright{r1}
  \fmf{fermion}{l1,r1}
 \end{fmfgraph*}
}},\hspace{5mm}
F_{f}(k)=
\parbox{20mm}{\unitlength=1mm\fmfframe(1,1)(1,1){
\begin{fmfgraph*}(15,1)\fmfpen{thin}
\fmfcmd{%
   style_def gap expr p=
     cdraw p;
     cfill (harrow(reverse p, .5));
     cfill (harrow(p,.5));
   enddef;
   style_def gap_hc expr p=
     cdraw p;
     cfill (tarrow(reverse p, .55));
     cfill (tarrow(p,.55));
   enddef;
     } 
\fmfleft{l1}
 \fmfright{r1}
  \fmf{gap}{l1,r1}
 \end{fmfgraph*}
}}\nonumber\\[5mm]
G_{\sigma}(k)&=&
\parbox{20mm}{\unitlength=1mm\fmfframe(1,1)(1,1){
\begin{fmfgraph*}(15,1)\fmfpen{thin} 
\fmfleft{l1}
 \fmfright{r1}
  \fmf{dashes}{l1,r1}
 \end{fmfgraph*}
}},\hspace{5mm}
G_{\pi}(k)=
\parbox{20mm}{\unitlength=1mm\fmfframe(1,1)(1,1){
\begin{fmfgraph*}(15,1)\fmfpen{thin} 
\fmfleft{l1}
 \fmfright{r1}
  \fmf{photon}{l1,r1}
 \end{fmfgraph*}
}}\nonumber\\[3mm]
g_{\sigma}&=&
\parbox{20mm}{\unitlength=1mm\fmfframe(2,2)(1,1){
\begin{fmfgraph*}(15,15)\fmfpen{thin}
\fmfleft{l1}
\fmfrightn{r}{2}
\fmf{dashes}{l1,G1}
\fmfpolyn{hatched,tension=0.8}{G}{3}
\fmf{fermion}{G2,r1}
\fmf{fermion}{G3,r2}
 \end{fmfgraph*}
}},\hspace{5mm}
g_{\pi}\hspace{4mm}=
\parbox{20mm}{\unitlength=1mm\fmfframe(2,2)(1,1){
\begin{fmfgraph*}(15,15)\fmfpen{thin}
\fmfleft{l1}
\fmfrightn{r}{2}
\fmf{photon}{l1,G1}
\fmfpolyn{filled=30,tension=0.8}{G}{3}
\fmf{fermion}{G2,r1}
\fmf{fermion}{G3,r2}
 \end{fmfgraph*}
}}\nonumber\\[3mm]
\gamma_{\sigma\pi^{2}}&=&
\parbox{20mm}{\unitlength=1mm\fmfframe(2,2)(1,1){
\begin{fmfgraph*}(15,15)\fmfpen{thin}
\fmfleft{l1}
\fmfrightn{r}{2}
\fmf{dashes}{l1,G1}
\fmfpolyn{full,tension=0.8}{G}{3}
\fmf{photon}{G2,r1}
\fmf{photon}{G3,r2}
 \end{fmfgraph*}
}},\hspace{5mm}
\gamma_{\sigma^{3}}\hspace{3mm}=
\parbox{20mm}{\unitlength=1mm\fmfframe(2,2)(1,1){
\begin{fmfgraph*}(15,15)\fmfpen{thin}
\fmfleft{l1}
\fmfrightn{r}{2}
\fmf{dashes}{l1,G1}
\fmfpolyn{full,tension=0.8}{G}{3}
\fmf{dashes}{G2,r1}
\fmf{dashes}{G3,r2}
 \end{fmfgraph*}
}}\nonumber\\[3mm]
\gamma_{\sigma^{4}}&=&
\parbox{20mm}{\unitlength=1mm\fmfframe(2,2)(1,1){
\begin{fmfgraph*}(15,15)\fmfpen{thin}
\fmfleftn{l}{2}
\fmfrightn{r}{2}
\fmf{dashes}{l2,G1}
\fmf{dashes}{l1,G2}
\fmfpolyn{full,tension=1.5}{G}{4}
\fmf{dashes}{r1,G3}
\fmf{dashes}{r2,G4}
 \end{fmfgraph*}
}},\hspace{5mm}
\gamma_{\sigma^{2}\pi^{2}}\hspace{1mm}=
\parbox{20mm}{\unitlength=1mm\fmfframe(2,2)(1,1){
\begin{fmfgraph*}(15,15)\fmfpen{thin}
\fmfleftn{l}{2}
\fmfrightn{r}{2}
\fmf{dashes}{l2,G1}
\fmf{dashes}{l1,G2}
\fmfpolyn{shaded,tension=1.5}{G}{4}
\fmf{photon}{r1,G3}
\fmf{photon}{r2,G4}
 \end{fmfgraph*}
}}\nonumber\\[3mm]
\gamma_{\pi^{4}}\hspace{1mm}&=&
\parbox{20mm}{\unitlength=1mm\fmfframe(2,2)(1,1){
\begin{fmfgraph*}(15,15)\fmfpen{thin}
\fmfleftn{l}{2}
\fmfrightn{r}{2}
\fmf{photon}{l2,G1}
\fmf{photon}{l1,G2}
\fmfpolyn{full,tension=1.5}{G}{4}
\fmf{photon}{r1,G3}
\fmf{photon}{r2,G4}
 \end{fmfgraph*}
}}\nonumber\\[-5mm]
\nonumber
\end{eqnarray}
\end{fmffile}
\caption{Diagrammatic constituents of our truncation in 
the symmetry-broken regime as specified in Sec. 
\ref{subsubsec:trunc_ssb}.}
\end{figure}

Let us now discuss terms involving fermions in the case of
symmetry breaking.
In addition to the normal quadratic fermionic term $\Gam_{\psib\psi}$ 
defined as before, see Eq.~(\ref{eq:Gam_psi_psi}), the
anomalous term
\begin{equation}
 \Gam_{\psi\psi} = \int_k \left(
 \Delta \, \psib_{-k\down} \psib_{k\up} + 
 \Delta^* \psi_{k\up} \psi_{-k\down} \right)
\end{equation}
is generated in the symmetry-broken regime, where $|\Delta|$ is 
a cutoff-dependent energy gap. 
The phase of $\Delta$ is inherited from the phase of 
$\alf$ while its modulus is generally different, 
due to fluctuations.
Since we have chosen $\alf$ real and positive, $\Delta$
is real and positive, too.
The normal and anomalous fermionic propagators $G_f$ and $F_f$ 
corresponding to $\Gam_{\psib\psi}$ and $\Gam_{\psi\psi}$ have 
the standard mean-field form as in Eqs.~(\ref{eq:Gf}) and 
(\ref{eq:Ff}), with $E_{\bk} = (\xi_{\bk}^2 + |\Delta|^2 )^{1/2}$, 
but now $\Delta$ is not equal to $\alf$.

In addition to the interaction between fermions and bosons 
of the form Eq.~(\ref{eq:g_sym}), an anomalous term of the 
form 
\begin{equation}
 \Gam_{\psi^2\phi} = 
  \tilde{g} \int_{k,q} \left(
  \psib_{-k+\frac{q}{2} \down} \psib_{k+\frac{q}{2} \up} \,
  \phi^*_{q} +
  \psi_{k+\frac{q}{2} \up} \psi_{-k+\frac{q}{2}\down} \,
  \phi_q \right)
\end{equation} 
is generated in the symmetry-broken regime.
Inserting the decomposition of $\phi$ in longitudinal and 
transverse fields into the normal and anomalous interaction
terms, we obtain
\begin{eqnarray}
 \Gam_{\psi^2\sg} &=& 
  g_{\sigma} \int_{k,q} \left( 
  \psib_{-k+\frac{q}{2} \down} \psib_{k+\frac{q}{2} \up} \,
  \sg_{q} +
  \psi_{k+\frac{q}{2} \up} \psi_{-k+\frac{q}{2}\down} \,
  \sg_{-q} \right) \, , \quad \\
 \Gam_{\psi^2\pi} &=& 
  ig_{\pi} \int_{k,q} \left( 
  \psib_{-k+\frac{q}{2} \down} \psib_{k+\frac{q}{2} \up} \,
  \pi_{q} - 
  \psi_{k+\frac{q}{2} \up} \psi_{-k+\frac{q}{2}\down} \,
  \pi_{-q} \right) \, ,
\end{eqnarray} 
where $g_{\sigma}=g+\tilde{g}$ and $g_{\pi}=g-\tilde{g}$. 
Fermions couple with different strength to the $\sigma$- 
and $\pi$-field, respectively.

A diagrammatic represention of the various progagators and 
interaction vertices in the symmetry-broken regime is shown
in Fig.~2.
We finally note that in a previously reported attempt\cite{krippa07} 
to truncate the fRG flow in a fermionic superfluid
the gap $\Delta$ was not taken into account in the fermionic 
propagator, and no distinction between longitudinal and 
transverse fields was made for the bosonic Z-factors in
the symmetry-broken regime.

\subsection{Approximate flow equations}
\label{subsec:flow_equations}

Inserting the above ansatz for the truncated effective action 
into the exact flow equation and comparing coefficients 
yields a set of coupled flow equations for the cutoff 
dependent parameters.
The various contributions can be conveniently represented 
by Feynman diagrams.
The prefactors and signs in the flow equations could be 
extracted from the expansion of the exact flow equation,
Eq.~(\ref{eq:flow_expansion}). However, in practice we determine 
them by comparison to a conventional perturbation expansion.

All contributions to our flow equations correspond to one-loop 
diagrams with only one momentum and frequency integration, 
as dictated by the structure of the exact flow equation in the 
form (\ref{eq:flow_expansion}). 
One of the propagators in the loop is a bosonic or fermionic
component of the single-scale propagator $\bG'_R$, the others
(if any) are components of $\bG_R$.
 
For a sharp frequency cutoff the frequency variable running 
around the loop is pinned by $\bG'_R(k_0)$ to $k_0 = \pm\Lam_b$
or $k_0 = \pm\Lam_f$. 
Hence the frequency integral can be performed analytically.
The problem that the integrand contains also step functions
$\chi_s^{\Lam}(k_0) = \Theta(|k_0|-\Lam_s)$ can be treated
by using the identity
$\int dx \, \delta(x-x_0) \, f[x,\Theta(x-x_0)] = 
\int_0^1 du \, f(x_0,u)$,
which is valid for any smooth function $f$.

More specifically, in the present case the one-loop diagrams 
are evaluated for vanishing external frequencies, such that
all internal propagators carry the same frequency.
In loops involving only either only bosonic or only fermionic 
propagators, one can use the identity
\begin{equation}
 n \int dk_0 \, \bG'_{sR}(k_0) \, \bA \, 
 [\bG_{sR}(k_0) \, \bA]^{n-1} = \Lam'_s
 \sum_{k_0 = \pm\Lam_s} [\bG_s(k_0) \, \bA]^n \, ,
\label{eq:cutoff_identity}
\end{equation}
valid for any matrix $\bA$,
to replace the frequency integration by a frequency sum
over $\pm\Lam_s$ while replacing all the propagators in the
loop by $\bG_s$. The factor $n$ corresponds to the $n$ possible
choices of positioning $\bG_{sR}'$ in a loop with $n$ lines,
and $\Lam_s' = \partial\Lam_s/\partial\Lam$.
For $\Lam_b = \Lam_f$ the above formula holds also for the
superpropagator $\bG_R$, such that is applies also to loops
with mixed products of bosonic and fermionic propagators.
For $\Lam_b \neq \Lam_f$, mixed loops contribute only if the
single-scale propagator is associated with the larger cutoff.
For example, for $\Lam_b > \Lam_f$, the single-scale propagator
has to be bosonic, since $\bG_b$ vanishes at $|k_0| = \Lam_f$.
On the other hand, for $|k_0| = \Lam_b$ one has 
$\bG_{fR}(k_0) = \bG_f(k_0)$, and for the integration of the
bosonic factors in the loop one can again use Eq.~(\ref{eq:cutoff_identity}).

For loop integrals involving the frequency sum over $\pm\Lam_s$
and the momentum integral over the Brillouin zone we use the
short-hand notation
\begin{equation}
 \int_{k|\Lam_s} = \frac{\Lam_s'}{2\pi} \sum_{k_0 = \pm\Lam_s}
 \int \frac{d^dk}{(2\pi)^d} \; ,
\end{equation}
where $\Lam_s' = \partial\Lam_s/\partial\Lam$.

\subsubsection{Symmetric regime}
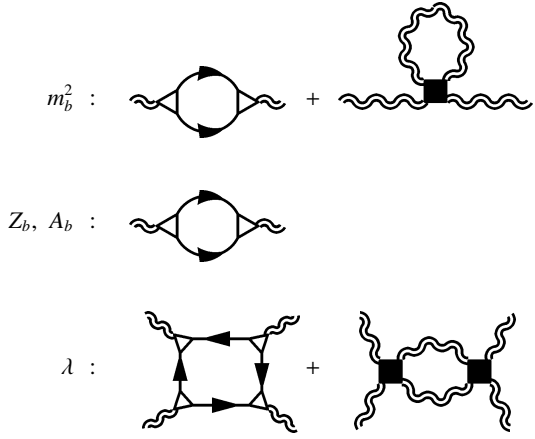
\begin{figure}
\begin{fmffile}{20080227_SYM_flow_1}
\begin{eqnarray}
m_{b}^{2}&:&
\parbox{25mm}{\unitlength=1mm\fmfframe(2,2)(1,1){
\begin{fmfgraph*}(20,20)\fmfpen{thin} 
 \fmfleft{l1}
 \fmfright{r1}
 \fmfpolyn{empty,tension=0.3}{G}{3}
 \fmfpolyn{empty,tension=0.3}{K}{3}
  \fmf{dbl_wiggly}{l1,G1}
 \fmf{fermion,tension=0.2,right=0.6}{G2,K3}
 \fmf{fermion,tension=0.2,left=0.6}{G3,K2}
 \fmf{dbl_wiggly}{K1,r1}
 \end{fmfgraph*}
}}+
\parbox{25mm}{\unitlength=1mm\fmfframe(2,2)(1,1){
\begin{fmfgraph*}(25,25)\fmfpen{thin} 
 \fmfleft{l1}
 \fmfright{r1}
 \fmftop{v1}
 \fmfpolyn{full}{G}{4}
 \fmf{dbl_wiggly,straight}{l1,G4}
 \fmf{dbl_wiggly,straight}{G1,r1}
 \fmffreeze
\fmf{dbl_wiggly,tension=0.1,right=0.7}{G2,v1}
\fmf{dbl_wiggly,tension=0.1,right=0.7}{v1,G3}
\end{fmfgraph*}
}}\nonumber\\[-8mm]
Z_{b},\,\,A_{b}&:&
\parbox{25mm}{\unitlength=1mm\fmfframe(2,2)(1,1){
\begin{fmfgraph*}(20,15)\fmfpen{thin} 
 \fmfleft{l1}
 \fmfright{r1}
 \fmfpolyn{empty,tension=0.3}{G}{3}
 \fmfpolyn{empty,tension=0.3}{K}{3}
  \fmf{dbl_wiggly}{l1,G1}
 \fmf{fermion,tension=0.2,right=0.6}{G2,K3}
 \fmf{fermion,tension=0.2,left=0.6}{G3,K2}
 \fmf{dbl_wiggly}{K1,r1}
 \end{fmfgraph*}
}}\nonumber\\
\lambda&:&
\parbox{25mm}{\unitlength=1mm\fmfframe(2,2)(1,1){
\begin{fmfgraph*}(25,15)
\fmfpen{thin} 
\fmfleftn{l}{2}\fmfrightn{r}{2}
 \fmfpolyn{empty,tension=0.8}{OL}{3}
 \fmfpolyn{empty,tension=0.8}{OR}{3}
 \fmfpolyn{empty,tension=0.8}{UR}{3}
 \fmfpolyn{empty,tension=0.8}{UL}{3}
   \fmf{dbl_wiggly}{l1,OL1}
   \fmf{dbl_wiggly}{l2,UL1}
   \fmf{dbl_wiggly}{OR1,r1}
   \fmf{dbl_wiggly}{UR1,r2}
  \fmf{fermion,straight,tension=0.5}{OL2,OR3}
  \fmf{fermion,straight,tension=0.5}{UR3,OR2}
  \fmf{fermion,straight,tension=0.5}{UR2,UL3}
  \fmf{fermion,straight,tension=0.5}{OL3,UL2}
 \end{fmfgraph*}
}}+
\parbox{25mm}{\unitlength=1mm\fmfframe(2,2)(1,1){
\begin{fmfgraph*}(25,15)
\fmfpen{thin} 
\fmfleftn{l}{2}\fmfrightn{r}{2}
\fmfrpolyn{full}{G}{4}
\fmfpolyn{full}{K}{4}
\fmf{dbl_wiggly}{l1,G1}\fmf{dbl_wiggly}{l2,G2}
\fmf{dbl_wiggly}{K1,r1}\fmf{dbl_wiggly}{K2,r2}
\fmf{dbl_wiggly,left=.5,tension=.3}{G3,K3}
\fmf{dbl_wiggly,right=.5,tension=.3}{G4,K4}
\end{fmfgraph*}
}}\nonumber
\end{eqnarray}
\end{fmffile}
\caption{Feynman diagrams representing the flow equations in the 
symmetric regime.}
\label{fig:flow_sym}
\end{figure}
Here we choose $\Lam_b = \Lam_f = \Lam$. One is in principle free 
to choose the fermionic and bosonic cutoff independently. We  have 
checked that the concrete choice of $\Lam_b$ and $\Lam_f$ in the symmetric 
regime does not change 
the final results for $\Lambda\rightarrow 0$ much. 
The Feynman diagrams contributing to the flow in the symmetric
regime are shown in Fig.~\ref{fig:flow_sym}.

The flow of the bosonic mass is given by the bosonic
self-energy at vanishing external momentum and frequency,
that is,
\begin{equation}
 \partial_{\Lam} \frac{m_b^2}{2} = 
 g^2 \int_{k|\Lam} G_f(k) \, G_f(-k) + 
 \frac{\lam}{2} \int_{q|\Lam} G_b(q) \; .
\label{eq:mass_sym}
\end{equation}
The fermionic contribution to $\partial_{\Lam} m_b^2$ is 
positive, leading to a reduction of $m_b^2$ upon decreasing 
$\Lam$, while the bosonic fluctuation term is negative 
(since $G_b(q) < 0$).
The flow of $Z_b$ is obtained from the second frequency
derivative of the bosonic self-energy as
\begin{equation}
 \partial_{\Lam} Z_b = 
 g^2 \int_{k|\Lam} \left. 
 \partial_{q_0}^2 G_f(k+q) \, G_f(-k) \right|_{q=0}
 \; .
\end{equation}
Similarly, the flow of $A_b$ is obtained from a second
momentum derivative of the bosonic self-energy:
\begin{equation}
 \partial_{\Lam} A_b = 
 g^2 \int_{k|\Lam} \left. 
 \partial_{\mathbf{q}}^2 G_f(k+q) \, G_f(-k) \right|_{q=0} \; ,
\end{equation}
where $\partial_{\mathbf{q}}^2=\frac{1}{d}\sum_{i=1}^{d}\partial_{q_i}^2$. 
Since the bosonic self-energy is isotropic in $\bq$ to
leading (quadratic) order in $\bq$, the results do not depend much on 
the direction in which the momentum derivative is taken.
The bosonic tadpole diagram in Fig.~3 contributes only to
$m_b$, not to $Z_b$ and $A_b$, since it yields a momentum 
and frequency independent contribution to the self-energy.
Finally, the flow of the $|\phi|^4$ coupling is given by
\begin{equation}
 \partial_{\Lam} \lam =
 - 4 g^4 \int_{k|\Lam} [G_f(k)]^2  [G_f(-k)]^2
 + \frac{5}{4} \lam^2 \int_{q|\Lam} [G_b(q)]^2 \; .
\label{eq:lambda}
\end{equation}

Within the truncation of the effective action described
in Sec.~IV.B there is no contribution to the flow of the
interaction between fermions and bosons in the symmetric
phase. The coupling $g$ remains therefore invariant.

\subsubsection{Symmetry broken regime}

In the limit $\Lam \to 0$ we are forced to choose 
$\Lam_f \ll \Lam_b$ to avoid an artificial strong coupling
problem, as will become clear below. We therefore choose
$\Lam_f < \Lam_b$ in the entire symmetry broken regime, 
which implies that the frequencies in mixed loops with
bosonic and fermionic propagators are pinned at the
bosonic cutoff. The precise choice of the cutoffs
will be specified later.
\begin{figure}
\begin{fmffile}{20080227_ferm_self_1}
\begin{eqnarray}
\Gamma_{\sigma}^{(1)}&:&
\parbox{15mm}{\unitlength=1mm\fmfframe(2,2)(1,1){
\begin{fmfgraph*}(15,30)\fmfpen{thin} 
\fmfleft{i1}\fmfright{o1}
\fmftop{t1}
\fmfpolyn{hatched,tension=0.2}{T}{3}
\fmfpolyn{phantom,tension=100.}{B}{3}
\fmf{phantom,straight,tension=1.2}{i1,B1}
\fmf{phantom,straight,tension=1.2}{o1,B2}
\fmf{dashes,tension=0.6}{B3,T3}
\fmf{fermion,tension=0.2,right=0.8}{T1,t1}\fmf{fermion,tension=0.2,left=0.8}{T2,t1}
\end{fmfgraph*}
}}+
\parbox{15mm}{\unitlength=1mm\fmfframe(2,2)(1,1){
\begin{fmfgraph*}(15,30)\fmfpen{thin} 
\fmfleft{i1}\fmfright{o1}
\fmftop{t1}
\fmfpolyn{full,tension=0.2}{T}{3}
\fmfpolyn{phantom,tension=100.}{B}{3}
\fmf{phantom,straight,tension=1.2}{i1,B1}
\fmf{phantom,straight,tension=1.2}{o1,B2}
\fmf{dashes,tension=0.6}{B3,T3}
\fmf{dashes,tension=0.2,right=0.8}{T1,t1}\fmf{dashes,tension=0.2,left=0.8}{T2,t1}
\end{fmfgraph*}
}}
+
\parbox{15mm}{\unitlength=1mm\fmfframe(2,2)(1,1){
\begin{fmfgraph*}(15,30)\fmfpen{thin} 
\fmfleft{i1}\fmfright{o1}
\fmftop{t1}
\fmfpolyn{full,tension=0.2}{T}{3}
\fmfpolyn{phantom,tension=100.}{B}{3}
\fmf{phantom,straight,tension=1.2}{i1,B1}
\fmf{phantom,straight,tension=1.2}{o1,B2}
\fmf{dashes,tension=0.6}{B3,T3}
\fmf{photon,tension=0.2,right=0.8}{T1,t1}\fmf{photon,tension=0.2,right=0.8}{t1,T2}
\end{fmfgraph*}
}}
\nonumber\\[-15mm]
\Delta&:& 
\parbox{25mm}{\unitlength=1mm\fmfframe(2,2)(1,1){
\begin{fmfgraph*}(22,15)\fmfpen{thin} 
\fmfcmd{%
   style_def gap expr p=
     cdraw p;
     cfill (harrow(reverse p, .5));
     cfill (harrow(p,.5));
   enddef;
   style_def gap_hc expr p=
     cdraw p;
     cfill (tarrow(reverse p, .55));
     cfill (tarrow(p,.55));
   enddef;
     }
\fmfleft{l1}
 \fmfright{r1}
 \fmfpolyn{hatched,tension=0.4}{G}{3}
 \fmfpolyn{hatched,tension=0.4}{K}{3}
  \fmf{fermion}{l1,G1}
   \fmf{gap_hc,straight,tension=0.5,left=0.}{K3,G2}
  \fmf{dashes,tension=0.,left=0.7}{G3,K2}
  \fmf{fermion}{r1,K1}
 \end{fmfgraph*}
}} -
\parbox{25mm}{\unitlength=1mm\fmfframe(2,2)(1,1){
\begin{fmfgraph*}(22,15)\fmfpen{thin} 
\fmfleft{l1}
 \fmfright{r1}
 \fmfpolyn{filled=30,tension=0.4}{G}{3}
 \fmfpolyn{filled=30,tension=0.4}{K}{3}
  \fmf{fermion}{l1,G1}
   \fmf{gap_hc,straight,tension=0.5,left=0.}{K3,G2}
  \fmf{photon,tension=0.,right=0.7}{K2,G3}
  \fmf{fermion}{r1,K1}
 \end{fmfgraph*}
}}\nonumber\\[-12mm]
\nonumber
\end{eqnarray}
\end{fmffile}
\caption{Contributions to the bosonic 1-point vertex and fermion gap below $\Lambda_{c}$.}
\label{fig:fermionic_self}
\end{figure}
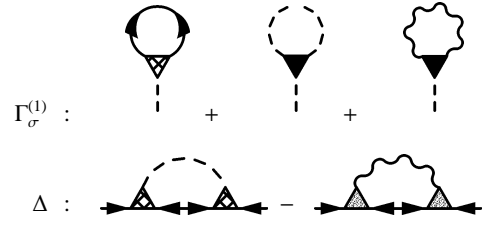

We first derive the flow equation for the minimum of the 
bosonic potential $\alf$, which is derived from the 
condition that the bosonic 1-point vertex $\Gam_{\sg}^{(1)}$
be zero for all $\Lam$.
The flow equation for $\Gam_{\sg}^{(1)}$ reads
\begin{equation}
 \partial_{\Lam} \Gam_{\sg}^{(1)} =
 m_{\sg}^2 \partial_{\Lam} \alf 
 + 2g_{\sg} \int_{k|\Lam_f} \! F_f(k)
 + \frac{\lam\alf}{2} \int_{q|\Lam_b} 
 [3G_{\sg}(q) + G_{\pi}(q)] \; . 
\end{equation}
The various contributions are represented diagrammatically
in Fig.~\ref{fig:fermionic_self}.
The first term is due to the cutoff-dependence of the
expansion point around which the effective action is 
expanded in powers of the fields.
The condition $\partial_{\Lam} \Gam_{\sg}^{(1)} = 0$ yields
\begin{equation}
 \partial_{\Lam} \alf =
 - \frac{2g_{\sg}}{m_{\sg}^2} \int_{k|\Lam_f} F_f(k) -
 \frac{1}{2\alf} \int_{q|\Lam_b} [3G_{\sg}(q) + G_{\pi}(q)] \; .
\label{eq:alpha_ssb}
\end{equation}
We have used Eq.~(\ref{eq:mSig_condition}) to simplify the last term.
The fermionic contribution to $\partial_{\Lam} \alf$ is 
negative, leading to an increase of $\alf$ upon decreasing 
$\Lam$, while the bosonic fluctuation term is positive and 
therefore reduces $\alf$. The behavior of Eq. (\ref{eq:alpha_ssb}) in the 
vicinity of the critical scale, $\Lambda\lesssim\Lambda_{c}$, when $\alpha$ 
and $m_{\sg}^2$ are small is 
shown below in Sec. \ref{subsec:critflow}.

The flow of $\Delta$ is obtained from the flow of the 
anomalous component of the fermionic self-energy as
\begin{equation}
 \partial_{\Lam} \Delta = 
 g_{\sg} \, \partial_{\Lam} \alf
 - \! \int_{q|\Lam_b} \! \left. F_f(q-k) \, 
 [g^2_{\sg}G_{\sg}(q) - g^2_{\pi}G_{\pi}(q)] \right|_{k=(0,\bk_F)} \; .
\label{eq:gap_ssb}
\end{equation}
The first term, due to the cutoff-dependence of the expansion
point for the effective action, links the flow of the fermionic 
gap to the flow of the bosonic order parameter.
The second term captures a correction to the relation 
between $\alf$ and $\Delta$ due to bosonic fluctuations as illustrated 
in Fig.~\ref{fig:fermionic_self}.
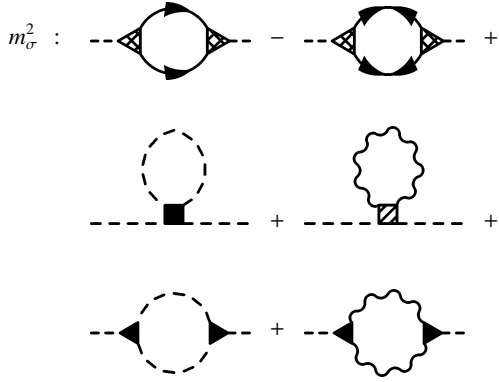
\begin{figure}
\begin{fmffile}{bos_self_48}
\begin{eqnarray}
m^{2}_{\sigma}&:&
\parbox{25mm}{\unitlength=1mm\fmfframe(2,2)(1,1){
\begin{fmfgraph*}(22,25)\fmfpen{thin} 
 \fmfleft{l1}
 \fmfright{r1}
 \fmfpolyn{hatched,tension=0.3}{G}{3}
 \fmfpolyn{hatched,tension=0.3}{K}{3}
  \fmf{dashes}{l1,G1}
 \fmf{fermion,tension=0.2,right=0.6}{G2,K3}
 \fmf{fermion,tension=0.2,left=0.6}{G3,K2}
 \fmf{dashes}{K1,r1}
 \end{fmfgraph*}
}}
-
\parbox{25mm}{\unitlength=1mm
\fmfframe(2,2)(1,1){
\begin{fmfgraph*}(22,25)
\fmfcmd{%
   style_def gap expr p=
     cdraw p;
     cfill (harrow(reverse p, .5));
     cfill (harrow(p,.5));
   enddef;
   style_def gap_hc expr p=
     cdraw p;
     cfill (tarrow(reverse p, .55));
     cfill (tarrow(p,.55));
   enddef;
     }
 \fmfpen{thin} 
 \fmfleft{l1}
 \fmfright{r1}
 \fmfpolyn{hatched,tension=0.3}{G}{3}
 \fmfpolyn{hatched,tension=0.3}{K}{3}
  \fmf{dashes}{l1,G1}
 \fmf{gap,tension=0.2,right=0.6}{G2,K3}
 \fmf{gap,tension=0.2,left=0.6}{G3,K2}
 \fmf{dashes}{K1,r1}
 \end{fmfgraph*}
}}+
\nonumber\\[-5mm]
&&
\parbox{25mm}{\unitlength=1mm\fmfframe(2,2)(1,1){
\begin{fmfgraph*}(22,25)\fmfpen{thin} 
 \fmfleft{l1}
 \fmfright{r1}
 \fmftop{v1}
 \fmfpolyn{full}{G}{4}
 \fmf{dashes,straight}{l1,G4}
 \fmf{dashes,straight}{G1,r1}
 \fmffreeze
\fmf{dashes,tension=0.1,right=0.7}{G2,v1}
\fmf{dashes,tension=0.1,right=0.7}{v1,G3}
\end{fmfgraph*}
}}
+
\parbox{25mm}{\unitlength=1mm\fmfframe(2,2)(1,1){
\begin{fmfgraph*}(22,25)\fmfpen{thin} 
 \fmfleft{l1}
 \fmfright{r1}
 \fmftop{v1}
 \fmfpolyn{shaded}{G}{4}
 \fmf{dashes,straight}{l1,G4}
 \fmf{dashes,straight}{G1,r1}
 \fmffreeze
\fmf{photon,tension=0.1,right=0.7}{G2,v1}
\fmf{photon,tension=0.1,right=0.7}{v1,G3}
\end{fmfgraph*}
}}+\nonumber\\[-15mm]
&& 
\parbox{25mm}{\unitlength=1mm\fmfframe(2,2)(1,1){
\begin{fmfgraph*}(22,25)\fmfpen{thin} 
 \fmfleft{l1}
 \fmfright{r1}
 \fmfpolyn{full,tension=0.4}{G}{3}
 \fmfpolyn{full,tension=0.4}{K}{3}
  \fmf{dashes}{l1,G1}
 \fmf{dashes,tension=0.2,right=0.8}{G2,K3}
 \fmf{dashes,tension=0.2,right=0.8}{K2,G3}
 \fmf{dashes}{K1,r1}
 \end{fmfgraph*}
}}
+
\parbox{25mm}{\unitlength=1mm\fmfframe(2,2)(1,1){
\begin{fmfgraph*}(22,25)\fmfpen{thin} 
 \fmfleft{l1}
 \fmfright{r1}
 \fmfpolyn{full,tension=0.4}{G}{3}
 \fmfpolyn{full,tension=0.4}{K}{3}
  \fmf{dashes}{l1,G1}
 \fmf{photon,tension=0.2,right=0.8}{G2,K3}
 \fmf{photon,tension=0.2,right=0.8}{K2,G3}
 \fmf{dashes}{K1,r1}
 \end{fmfgraph*}
}}
\nonumber\\[-15mm]\nonumber
\end{eqnarray}
\end{fmffile}
\caption{Diagrammatic representation of the contributions to the 
bosonic mass in the symmetry-broken regime.}
\label{fig:bosonic_self}
\end{figure}

The flow of the mass of the longitudinal order parameter 
fluctuations (cf.\ Fig.~\ref{fig:bosonic_self}) is obtained from the 
self-energy of the $\sg$ fields at zero momentum and frequency as
\begin{eqnarray}
 \partial_{\Lam} \frac{m_{\sg}^2}{2} &=&
 g^2_{\sg} \int_{k|\Lam_f}
 \left[ G_f(k) G_f(-k) - F_f(k) F_f(-k) \right] + 
 3 \frac{\lam \alf}{2} \, \partial_{\Lam}\alf \nonumber \\
 &+& \frac{\lam}{4} \int_{q|\Lam_b} 
 \left[ 3G_{\sg}(q) + G_{\pi}(q) \right] 
 \nonumber \\
 &+& \frac{(\lam\alf)^2}{2} \int_{q|\Lam_b} 
 \left[ 9G_{\sg}^2(q) + G_{\pi}^2(q) \right] \; .
\label{eq:mass_ssb}
\end{eqnarray}
The second term in this equation is due to a product of the 3-point
vertex $\gam_{\sg^3}$ and $\partial_{\Lam}\alf$ arising from the 
cutoff dependence of the expansion point for the effective action. 
The flow of $\lam$ can be computed from the flow of 
$m_{\sg}^2$ and $\alf$ via the relation Eq.~(\ref{eq:mSig_condition}), 
$\lam = m_{\sg}^2/|\alf|^2$.

The flow of $Z_{\sg}$ is obtained from the second frequency
derivative of the self-energy of the $\sg$ fields, which yields
\begin{eqnarray}
 \partial_{\Lam} Z_{\sg} &=& 
 g^2_{\sg} \! \int_{k|\Lam_f} \!\!\! \left. \partial_{q_0}^2 
 [G_f(k+q) \, G_f(-k) - F_f(k+q) \, F_f(-k)] 
 \right|_{q=0} \nonumber \\
 + && \hskip -5mm \frac{(\lam\alf)^2}{2} \! \int_{k|\Lam_b} \!\!\!  
 \left. \partial_{q_0}^2 
 [9G_{\sg}(k+q) \, G_{\sg}(k) 
 + G_{\pi}(k+q) \, G_{\pi}(k)] \right|_{q=0} \; .
 \nonumber \\ &&
\end{eqnarray}
The flow of $A_{\sg}$ is given by the same equation with 
$\partial_{q_0}^2$ replaced by $\partial_{\mathbf{q}}^2$.

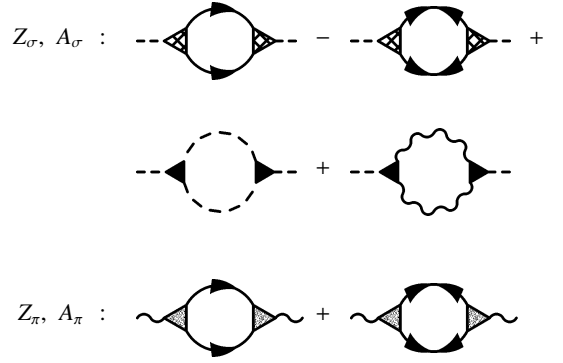
\begin{figure}
\begin{fmffile}{Z_factors_9}
\begin{eqnarray}
Z_{\sigma},\,\,A_{\sigma}&:&
\parbox{25mm}{\unitlength=1mm\fmfframe(2,2)(1,1){
\begin{fmfgraph*}(22,25)\fmfpen{thin}
 \fmfleft{l1}
 \fmfright{r1}
 \fmfpolyn{hatched,tension=0.3}{G}{3}
 \fmfpolyn{hatched,tension=0.3}{K}{3}
  \fmf{dashes}{l1,G1}
 \fmf{fermion,tension=0.2,right=0.6}{G2,K3}
 \fmf{fermion,tension=0.2,left=0.6}{G3,K2}
 \fmf{dashes}{K1,r1}
 \end{fmfgraph*}
}}
-
\parbox{25mm}{\unitlength=1mm
\fmfframe(2,2)(1,1){
\begin{fmfgraph*}(22,25)
\fmfcmd{%
   style_def gap expr p=
     cdraw p;
     cfill (harrow(reverse p, .5));
     cfill (harrow(p,.5));
   enddef;
   style_def gap_hc expr p=
     cdraw p;
     cfill (tarrow(reverse p, .55));
     cfill (tarrow(p,.55));
   enddef;
     }
 \fmfpen{thin} 
 \fmfleft{l1}
 \fmfright{r1}
 \fmfpolyn{hatched,tension=0.3}{G}{3}
 \fmfpolyn{hatched,tension=0.3}{K}{3}
  \fmf{dashes}{l1,G1}
 \fmf{gap,tension=0.2,right=0.6}{G2,K3}
 \fmf{gap,tension=0.2,left=0.6}{G3,K2}
 \fmf{dashes}{K1,r1}
 \end{fmfgraph*}
}}+
\nonumber\\[-12mm]
&& 
\parbox{25mm}{\unitlength=1mm\fmfframe(2,2)(1,1){
\begin{fmfgraph*}(22,25)\fmfpen{thin} 
 \fmfleft{l1}
 \fmfright{r1}
 \fmfpolyn{full,tension=0.4}{G}{3}
 \fmfpolyn{full,tension=0.4}{K}{3}
  \fmf{dashes}{l1,G1}
 \fmf{dashes,tension=0.2,right=0.8}{G2,K3}
 \fmf{dashes,tension=0.2,right=0.8}{K2,G3}
 \fmf{dashes}{K1,r1}
 \end{fmfgraph*}
}}
+
\parbox{25mm}{\unitlength=1mm\fmfframe(2,2)(1,1){
\begin{fmfgraph*}(22,25)\fmfpen{thin} 
 \fmfleft{l1}
 \fmfright{r1}
 \fmfpolyn{full,tension=0.4}{G}{3}
 \fmfpolyn{full,tension=0.4}{K}{3}
  \fmf{dashes}{l1,G1}
 \fmf{photon,tension=0.2,right=0.8}{G2,K3}
 \fmf{photon,tension=0.2,right=0.8}{K2,G3}
 \fmf{dashes}{K1,r1}
 \end{fmfgraph*}
}}
\nonumber\\[-10mm]
Z_{\pi},\,\,A_{\pi}&:&
\parbox{25mm}{\unitlength=1mm\fmfframe(2,2)(1,1){
\begin{fmfgraph*}(22,25)\fmfpen{thin}
 \fmfleft{l1}
 \fmfright{r1}
 \fmfpolyn{filled=30,tension=0.3}{G}{3}
 \fmfpolyn{filled=30,tension=0.3}{K}{3}
  \fmf{photon}{l1,G1}
 \fmf{fermion,tension=0.2,right=0.6}{G2,K3}
 \fmf{fermion,tension=0.2,left=0.6}{G3,K2}
 \fmf{photon}{K1,r1}
 \end{fmfgraph*}
}}
+
\parbox{25mm}{\unitlength=1mm
\fmfframe(2,2)(1,1){
\begin{fmfgraph*}(22,25)
\fmfcmd{%
   style_def gap expr p=
     cdraw p;
     cfill (harrow(reverse p, .5));
     cfill (harrow(p,.5));
   enddef;
   style_def gap_hc expr p=
     cdraw p;
     cfill (tarrow(reverse p, .55));
     cfill (tarrow(p,.55));
   enddef;
     }
 \fmfpen{thin} 
 \fmfleft{l1}
 \fmfright{r1}
 \fmfpolyn{filled=30,tension=0.3}{G}{3}
 \fmfpolyn{filled=30,tension=0.3}{K}{3}
  \fmf{photon}{l1,G1}
 \fmf{gap,tension=0.2,right=0.6}{G2,K3}
 \fmf{gap,tension=0.2,left=0.6}{G3,K2}
 \fmf{photon}{K1,r1}
 \end{fmfgraph*}
}}\nonumber\\[-10mm]
\nonumber
\end{eqnarray}
\end{fmffile}
\caption{Diagrams renormalizing the bosonic 
Z- and A-factors for $\Lambda<\Lambda_{c}$.}
\label{fig:bosonic_Z}
\end{figure}

In the flow of $Z_{\pi}$ there are strong cancellations of
different terms originating from bosonic fluctuations.
These cancellations are related to Ward idenities which
guarantee that $Z_{\pi}$ remains finite such that the
Goldstone mode is not renormalized substantially.\cite{pistolesi04} 
Hence, we keep only the fermionic fluctuations, that is,
\begin{equation}
 \partial_{\Lam} Z_{\pi} = 
 g^2_{\pi} \! \int_{k|\Lam_f} \!\!\! \left. \partial_{q_0}^2 
 [G_f(k+q) \, G_f(-k) + F_f(k+q) \, F_f(-k)] 
 \right|_{q=0} \; .
\end{equation}
For the flow of $A_{\pi}$ we obtain the same equation with 
$\partial_{q_0}^2$ replaced by $\partial_{\mathbf{q}}^2$.
The terms contributing to the flow of the $Z$- and $A$-factors 
are illustrated diagrammatically in Fig.~\ref{fig:bosonic_Z}.

\begin{figure}
\begin{fmffile}{20080214_vertex_19}
\begin{eqnarray}
g_{\sigma}&:&
\parbox{30mm}{\unitlength=1mm\fmfframe(2,2)(1,1){
\begin{fmfgraph*}(25,25)
\fmfcmd{%
   style_def gap expr p=
     cdraw p;
     cfill (harrow(reverse p, .5));
     cfill (harrow(p,.5));
   enddef;
   style_def gap_hc expr p=
     cdraw p;
     cfill (tarrow(reverse p, .55));
     cfill (tarrow(p,.55));
   enddef;
     }
\fmfpen{thin} 
\fmfleftn{l}{2}\fmfright{r}
\fmfrpolyn{hatched,tension=0.65}{T}{3}
\fmfrpolyn{hatched,tension=0.65}{B}{3}
\fmfpolyn{hatched, tension=0.65}{R}{3}
\fmf{fermion}{l1,T2}\fmf{fermion}{l2,B2}
\fmf{dashes,tension=.6}{T3,B1}
\fmf{gap_hc,left=0.,tension=.5}{T1,R3}
\fmf{gap_hc,right=0.,tension=.5}{B3,R2}
\fmf{dashes}{R1,r}
\end{fmfgraph*}
}}
-
\hspace{-2mm}
\parbox{30mm}{\unitlength=1mm\fmfframe(2,2)(1,1){
\begin{fmfgraph*}(25,25)
\fmfcmd{%
   style_def gap expr p=
     cdraw p;
     cfill (harrow(reverse p, .5));
     cfill (harrow(p,.5));
   enddef;
   style_def gap_hc expr p=
     cdraw p;
     cfill (tarrow(reverse p, .55));
     cfill (tarrow(p,.55));
   enddef;
     }
\fmfpen{thin} 
\fmfleftn{l}{2}\fmfright{r}
\fmfrpolyn{filled=30,tension=0.65}{T}{3}
\fmfrpolyn{filled=30,tension=0.65}{B}{3}
\fmfpolyn{hatched, tension=0.65}{R}{3}
\fmf{fermion}{l1,T2}\fmf{fermion}{l2,B2}
\fmf{photon,tension=.5,left=0.}{B1,T3}
\fmf{gap_hc,left=0.,tension=.5}{T1,R3}
\fmf{gap_hc,right=0.,tension=.5}{B3,R2}
\fmf{dashes}{R1,r}
\end{fmfgraph*}
}}
\nonumber\\[0mm]
-
&&
\parbox{30mm}{\unitlength=1mm\fmfframe(2,2)(1,1){
\begin{fmfgraph*}(25,25)
\fmfpen{thin} 
\fmfleftn{l}{2}\fmfright{r}
\fmfrpolyn{hatched,tension=0.65}{T}{3}
\fmfrpolyn{hatched,tension=0.65}{B}{3}
\fmfpolyn{hatched, tension=0.65}{R}{3}
\fmf{fermion}{l1,T2}\fmf{fermion}{l2,B2}
\fmf{dashes,tension=.6}{T3,B1}
\fmf{fermion,left=0.,tension=.5}{R3,T1}
\fmf{fermion,right=0.,tension=.5}{R2,B3}
\fmf{dashes}{R1,r}
\end{fmfgraph*}
}}
+
\hspace{-2mm}
\parbox{30mm}{\unitlength=1mm\fmfframe(2,2)(1,1){
\begin{fmfgraph*}(25,25)
\fmfpen{thin} 
\fmfleftn{l}{2}\fmfright{r}
\fmfrpolyn{filled=30,tension=0.65}{T}{3}
\fmfrpolyn{filled=30,tension=0.65}{B}{3}
\fmfpolyn{hatched, tension=0.65}{R}{3}
\fmf{fermion}{l1,T2}\fmf{fermion}{l2,B2}
\fmf{photon,tension=.5,left=0.}{B1,T3}
\fmf{fermion,left=0.,tension=.5}{R3,T1}
\fmf{fermion,right=0.,tension=.5}{R2,B3}
\fmf{dashes}{R1,r}
\end{fmfgraph*}
}}
\nonumber\\[5mm]
g_{\pi}&:&
\parbox{30mm}{\unitlength=1mm\fmfframe(2,2)(1,1){
\begin{fmfgraph*}(25,25)
\fmfcmd{%
   style_def gap expr p=
     cdraw p;
     cfill (harrow(reverse p, .5));
     cfill (harrow(p,.5));
   enddef;
   style_def gap_hc expr p=
     cdraw p;
     cfill (tarrow(reverse p, .55));
     cfill (tarrow(p,.55));
   enddef;
     }
\fmfpen{thin} 
\fmfleftn{l}{2}\fmfright{r}
\fmfrpolyn{hatched,tension=0.65}{T}{3}
\fmfrpolyn{hatched,tension=0.65}{B}{3}
\fmfpolyn{filled=30, tension=0.65}{R}{3}
\fmf{fermion}{l1,T2}\fmf{fermion}{l2,B2}
\fmf{dashes,tension=.6}{T3,B1}
\fmf{gap_hc,left=0.,tension=.5}{T1,R3}
\fmf{gap_hc,right=0.,tension=.5}{B3,R2}
\fmf{photon}{R1,r}
\end{fmfgraph*}
}}
-
\hspace{-2mm}
\parbox{30mm}{\unitlength=1mm\fmfframe(2,2)(1,1){
\begin{fmfgraph*}(25,25)
\fmfcmd{%
   style_def gap expr p=
     cdraw p;
     cfill (harrow(reverse p, .5));
     cfill (harrow(p,.5));
   enddef;
   style_def gap_hc expr p=
     cdraw p;
     cfill (tarrow(reverse p, .55));
     cfill (tarrow(p,.55));
   enddef;
     }
\fmfpen{thin} 
\fmfleftn{l}{2}\fmfright{r}
\fmfrpolyn{filled=30,tension=0.65}{T}{3}
\fmfrpolyn{filled=30,tension=0.65}{B}{3}
\fmfpolyn{filled=30, tension=0.65}{R}{3}
\fmf{fermion}{l1,T2}\fmf{fermion}{l2,B2}
\fmf{photon,tension=.5,left=0.}{B1,T3}
\fmf{gap_hc,left=0.,tension=.5}{T1,R3}
\fmf{gap_hc,right=0.,tension=.5}{B3,R2}
\fmf{photon}{R1,r}
\end{fmfgraph*}
}}
\nonumber\\[0mm]
+
&&
\parbox{30mm}{\unitlength=1mm\fmfframe(2,2)(1,1){
\begin{fmfgraph*}(25,25)
\fmfpen{thin} 
\fmfleftn{l}{2}\fmfright{r}
\fmfrpolyn{hatched,tension=0.65}{T}{3}
\fmfrpolyn{hatched,tension=0.65}{B}{3}
\fmfpolyn{filled=30, tension=0.65}{R}{3}
\fmf{fermion}{l1,T2}\fmf{fermion}{l2,B2}
\fmf{dashes,tension=.6}{T3,B1}
\fmf{fermion,left=0.,tension=.5}{R3,T1}
\fmf{fermion,right=0.,tension=.5}{R2,B3}
\fmf{photon}{R1,r}
\end{fmfgraph*}
}}
-
\hspace{-2mm}
\parbox{30mm}{\unitlength=1mm\fmfframe(2,2)(1,1){
\begin{fmfgraph*}(25,25)
\fmfpen{thin} 
\fmfleftn{l}{2}\fmfright{r}
\fmfrpolyn{filled=30,tension=0.65}{T}{3}
\fmfrpolyn{filled=30,tension=0.65}{B}{3}
\fmfpolyn{filled=30, tension=0.65}{R}{3}
\fmf{fermion}{l1,T2}\fmf{fermion}{l2,B2}
\fmf{photon,tension=.5,left=0.}{B1,T3}
\fmf{fermion,left=0.,tension=.5}{R3,T1}
\fmf{fermion,right=0.,tension=.5}{R2,B3}
\fmf{photon}{R1,r}
\end{fmfgraph*}
}}\nonumber\\[-5mm]
\nonumber
\end{eqnarray}
\end{fmffile}
\caption{Fermion-boson vertex corrections below $\Lambda_{c}$.}
\label{fig:vertex_ssb}
\end{figure}
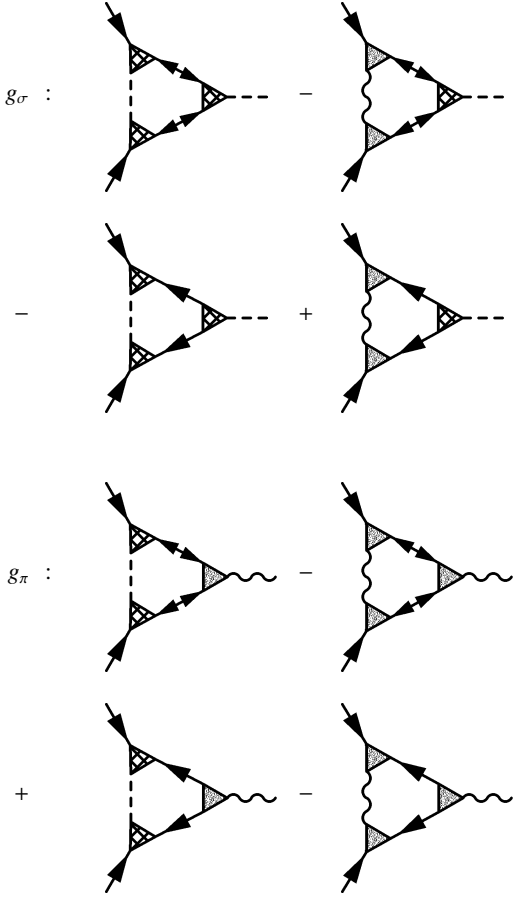

In the symmetry broken regime, there are also contributions 
to the flow of the interaction between fermions and bosons
due to vertex corrections with bosonic fluctuations
(see Fig.~\ref{fig:vertex_ssb}), yielding
\begin{eqnarray}
 \partial_{\Lam} g_{\sigma} = g_{\sigma} \int_{q|\Lam_b}
 && 
 \left[F^{2}_f(k-q) - \left|G_f(k-q)\right|^2\right]_{k=(0,\bk_F)} 
 \nonumber\\ 
 &&\left[g_{\sigma}^{2}G_{\sg}(q) - g_{\pi}^{2}G_{\pi}(q)\right]
 \; ,
\end{eqnarray}
and
\begin{eqnarray}
 \partial_{\Lam} g_{\pi} = g_{\pi} \int_{q|\Lam_b}
 && 
 \left[F^{2}_f(k-q) + \left|G_f(k-q)\right|^2\right]_{k=(0,\bk_F)}
 \nonumber\\ 
 && \left[g_{\sigma}^{2}G_{\sg}(q) - g_{\pi}^{2}G_{\pi}(q)\right]
 \; .
\label{eq:g_pi_ssb}
\end{eqnarray}
The right hand sides are dominated by the contribution from
the $\pi$ propagator, which tends to reduce $g_{\sigma}$, $g_{\pi}$  for 
decreasing $\Lam$.

\subsection{Relation to mean-field theory}

Before solving the flow equations derived above, we first
analyze what happens when contributions due to bosonic 
fluctuations are neglected, and relate the reduced set of
equations to the usual mean-field theory (Sec.~III).

In the absence of bosonic fluctuations, $g=g_{\sg}=g_{\pi}=1$.
Furthermore, the bosonic order parameter $\alf$ and the
fermionic gap $\Delta$ are identical: $\alf = \Delta$.
Since the bosonic cutoff is irrelevant here, we can 
choose $\Lam_f = \Lam$.

The flow equation for the bosonic mass in the symmetric 
regime, Eq.~(\ref{eq:mass_sym}), simplifies to
\begin{equation}
 \partial_{\Lam} \frac{m_b^2}{2} = 
 \int_{k|\Lam} G_f(k) \, G_f(-k) \; ,
\end{equation}
where $G_f(k) = G_{f0}(k) = (ik_0 - \xi_{\bk})^{-1}$.
This equation can be easily integrated, yielding
\begin{equation}
 \frac{m_b^2}{2} = \frac{1}{|U|} - 
 \int_{|k_0|> \Lam} \int_{\bk} 
 \frac{1}{k_0^2 + \xi_{\bk}^2} \; .
\end{equation}
$m_b$ vanishes at a critical scale $\Lam_c > 0$.

The flow equation for $\Delta$ (= $\alf$) in the symmetry
broken regime $\Lam < \Lam_c \,$, Eq.~(\ref{eq:gap_ssb}), is reduced to
\begin{eqnarray}
 \partial_{\Lam} \Delta =
 - \frac{2}{m_{\sg}^2} \int_{k|\Lam} F_f(k)  \; .
\label{eq:gap_mftflow}
\end{eqnarray}
It is complemented by the flow equation for the mass of
the $\sg$ field, Eq.~(\ref{eq:mass_ssb}), which becomes
\begin{eqnarray}
 \partial_{\Lam} \frac{m_{\sg}^2}{2} =
 \int_{k|\Lam}
 \left[ |G_f(k)|^2 - F_f^2(k) \right]
 + 3 \gam_{\sg^3} \, \partial_{\Lam}\Delta
\label{eq:mass_mftflow}
\end{eqnarray}
in the absence of bosonic fluctuations,
with $\gam_{\sg^3} = \lam\Delta/2 = m_{\sg}^2/(2\Delta)$.
The propagators $G_f$ and $F_f$ have the usual BCS form, 
as in Eqs.~(\ref{eq:Gf}) and (\ref{eq:Ff}).

A numerical solution of the coupled flow equations (\ref{eq:gap_mftflow}) 
and (\ref{eq:mass_mftflow}) yields a gap $\Delta$ which is a bit smaller than 
the BCS result obtained from the gap equation (\ref{eq:bcs_gap_eqn}).
The reason for this discrepancy is the relatively simple
quartic ansatz (\ref{eq:U_local}) for the bosonic potential. 
The complete bosonic potential is non-polynomial in 
$|\phi|^2$ even in mean-field theory.
Restricted to the zero momentum and frequency component
of $\phi$ it has the form \cite{popov87}
\begin{equation}
 U^{\rm MF}(\phi) = \frac{|\phi|^2}{|U|} - \int_k \ln 
 \frac{k_0^2 + \xi_{\bk}^2 + |\phi|^2}{k_0^2 + \xi_{\bk}^2}
 \; .
\end{equation}
The kernel of the 3-point vertex $\gam_{\sg^3}$ obtained 
from an expansion of this mean-field potential around a 
finite order parameter $\Delta$ reads
\begin{equation}
 \gam_{\sg^3} = - 2 \int_k \; \Big[
 \frac{1}{3} F_f^3(k) - F_f(k) \, |G_f(k)|^2 \Big]
\end{equation}
at zero frequencies and momenta.
Inserting this into (\ref{eq:mass_mftflow}), the flow of $m_{\sg}^2$ can
be written as a total derivative
\begin{equation}
 \partial_{\Lam} \frac{m_{\sg}^2}{2} =
 - \partial_{\Lam} \int_{|k_0| > \Lam} \int_{\bk}
 \left[ |G_f(k)|^2 - F_f^2(k) \right] \; ,
\label{eq:mass_mftflow_correct}
\end{equation}
where the $\Lam$-derivative on the right hand side acts
also on $\Delta$, generating the term proportional to
$\gam_{\sg^3}$.
Integrating this equation with the initial condition 
$m_{\sg} = 0$ at $\Lam=\Lam_c \,$, yields
\begin{equation}
 \frac{m_{\sg}^2}{2} = K(0) + L(0) - U^{-1} \; ,
\label{eq:mass_mft_correct}
\end{equation}
which is the correct mean-field result.
With $m_{\sg}$ given by (\ref{eq:mass_mft_correct}), the 
flow equation (\ref{eq:gap_mftflow}) yields the correct mean-field gap. 
The easiest way to see this, is to write the BCS gap
equation in the presence of a cutoff in the form
$1 = - U \int_{|k_0| > \Lam} \int_{\bk} \Delta^{-1} \,
 F_f(k)$, and take a derivative with respect to $\Lam$.

It is instructive to relate the above flow equations 
for $\Delta$ and $m_{\sg}$ to the flow equations for 
the BCS mean-field model obtained in a purely fermionic 
RG.\cite{salmhofer04}
For a sharp frequency cutoff, those flow equations 
have the form
\begin{eqnarray}
 \partial_{\Lam} \Delta &=&
 - (V+W) \int_k' F_f(k)  \; , \\
 \partial_{\Lam} (V \!+\! W) &=&
 (V \!+\! W)^2 \, \partial_{\Lam} \!\! 
 \int_{|k_0| > \Lam} \int_{\bk}
 \left[ |G_f(k)|^2 - F_f^2(k) \right] \, , \hskip 5mm
\end{eqnarray}
where $V$ is a normal two-fermion interaction in the Cooper 
channel, while $W$ is an anomalous interaction 
corresponding to annihilation (or creation) of four 
particles. With the identification $2/m_{\sg}^2 = V+W$ 
these equations are obviously equivalent to (\ref{eq:gap_mftflow})
and (\ref{eq:mass_mftflow_correct}).
The above flow equation for $V+W$ is obtained from a one-loop
truncation complemented by additional self-energy
insertions drawn from higher order diagrams with tadpoles.
\cite{salmhofer04,katanin04}
These additional terms correspond to the contractions with 
$\gam_{\sg^3}$ in the present bosonized RG.

\section{Results}

\subsection{Flow for $\Lam \lesssim \Lam_c$}
\label{subsec:critflow}

For $\Lam$ slightly below $\Lam_c$, the flow equations can
be expanded in the order parameter. To leading order, the
order parameter $\alf$ and the gap $\Delta$ are identical, 
$\alf = \Delta \,$. 
The fluctuation term in the flow equation (\ref{eq:gap_ssb}) for $\Delta$
is quadratic in $\alf$, and also the flow of $g_{\sigma}$ yields
only corrections beyond linear order to the relation between 
$\alf$ and $\Delta$.

Near $\Lam_c$, the flow equation (\ref{eq:alpha_ssb}) for $\alf$ can be
written as
\begin{equation}
 \partial_{\Lam} \alf^2 = 
 - \frac{4}{\lam} \, I 
 - \int_{q|\Lam_b} [3G_{\sg}(q) + G_{\pi}(q)] \; ,
\end{equation}
where $I = \int_{k|\Lam_f} (k_0^2 + \tilde\xi_{\bk}^2)^{-1}$, 
evaluated for $\Lam = \Lam_c$. Note that we have replaced
the ratio $m_{\sg}^2/\alf^2$ by $\lam$ in the first term
on the right hand side of the flow equation.
Integrating the flow equation one obtains
\begin{equation}
 \alf^2 = 
 \left[ \frac{4}{\lam} \, I +
 \int_{q|\Lam_b} [3G_{\sg}(q) + G_{\pi}(q)] \right]_{\Lam=\Lam_c} 
 (\Lam_c - \Lam)
\label{eq:alpha_L_c}
\end{equation}
for $\Lam \lesssim \Lam_c$. The order parameter $\alf$ and
the fermionic gap are thus proportional to
$(\Lam_c - \Lam)^{1/2}$ for $\Lam \lesssim \Lam_c$.

Inserting Eq.~(\ref{eq:alpha_L_c}) into the flow equation 
(\ref{eq:mass_ssb}) 
for $m_{\sg}^2$ and neglecting the last fluctuation term, 
which is of higher order in $\alf$, one obtains
\begin{equation}
 \partial_{\Lam} m_{\sg}^2 = 
 - 4 \, I
 - \lam \int_{q|\Lam_b} [3G_{\sg}(q) + G_{\pi}(q)] \; .
\end{equation}
This shows that the flow of $\alf$ and $m_{\sg}^2$ is 
indeed consistent with the relation 
$m_{\sg}^2 = \lam \alf^2$ following from the ansatz for
the bosonic potential.

\subsection{Infrared asymptotics}
\label{subsec:asymptotics}

In the infrared limit ($\Lam \to 0$) the key properties of the
flow can be extracted from the flow equations analytically.
The behavior of the bosonic sector depends strongly on the
dimensionality of the system. We consider dimensions $d \geq 2$,
focusing in particular on the two- and three-dimensional case.

The bosonic order parameter and the fermionic gap saturate at 
finite values in the limit $\Lam \to 0$. The fluctuation
corrections to $\partial_{\Lam} \alf$ and $\partial_{\Lam} \Delta$
involve the singular Goldstone propagator $G_{\pi}$ only linearly
and are therefore integrable in $d > 1$.
The fermion-boson interactions $g_{\sg}$ and $g_{\pi}$ also saturate. 
The finiteness of $Z_{\pi}$ and $A_{\pi}$ is guaranteed by Ward
identities.\cite{pistolesi04} In our truncation $Z_{\pi}$ and $A_{\pi}$
remain finite since only fermionic fluctuations contribute to
the Goldstone propagator. 
We choose $\Lam_b = \Lam$ in the following. The choice of $\Lam_f$ 
(as a function of $\Lam$) will be discussed and specified below.

\begin{figure}
\begin{fmffile}{goldstone_1}
\begin{eqnarray}
m^{2}_{\sigma},\,\,Z_{\sigma},\,\,A_{\sigma}&:&
\parbox{25mm}{\unitlength=1mm\fmfframe(2,2)(1,1){
\begin{fmfgraph*}(22,25)\fmfpen{thin} 
 \fmfleft{l1}
 \fmfright{r1}
 \fmfpolyn{full,tension=0.4}{G}{3}
 \fmfpolyn{full,tension=0.4}{K}{3}
  \fmf{dashes}{l1,G1}
 \fmf{photon,tension=0.2,right=0.8}{G2,K3}
 \fmf{photon,tension=0.2,right=0.8}{K2,G3}
 \fmf{dashes}{K1,r1}
 \end{fmfgraph*}
}}
\nonumber\\[-15mm]
\nonumber
\end{eqnarray}
\end{fmffile}
\caption{Goldstone fluctuations determining the infrared asymptotics 
for $\Lambda\rightarrow0$.}
\label{fig:goldstone}
\end{figure}
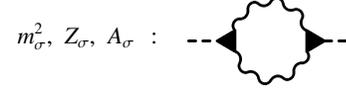

The flows of $m_{\sg}^2$, $\lam$, $Z_{\sg}$, and $A_{\sg}$ are 
dominated by terms quadratic in $G_{\pi}$ for $\Lam \to 0$, see
Fig.~\ref{fig:goldstone}.
Using $m_{\sg}^2 = \lam \alf^2$, we obtain the asymptotic flow
equation for $\lam$ from Eq.~(\ref{eq:mass_ssb}) in the simple 
form
\begin{equation}
 \partial_{\Lam} \lam = \lam^2 \int_{q|\Lam} G_{\pi}^2(q) \; .
\end{equation}
The above integral over $G_{\pi}^2$ is proportional to $\Lam^{d-4}$ 
for small $\Lam$ in dimensions $d < 4$, implying that $\lam$ 
scales to zero in $d \leq 3$.
In two dimensions one obtains $\int_{q|\Lam} G_{\pi}^2(q) = 
\frac{1}{4\pi^2 A_{\pi} Z_{\pi}} \Lam^{-2}$ for small $\Lam$, 
such that the rescaled variable $\tilde\lam = \lam/\Lam$ obeys
the flow equation
\begin{equation}
 \frac{d\tilde\lam}{d\log\Lam} = - \tilde\lam +
 \frac{\tilde\lam^2}{4\pi^2 A_{\pi} Z_{\pi}} \; ,
\end{equation}
which has a stable fixed point at 
$\tilde\lam^* = 4\pi^2 A_{\pi} Z_{\pi}$. Hence, the bosonic
self-interaction vanishes as
\begin{equation}
 \lam \to 4\pi^2 A_{\pi} Z_{\pi} \, \Lam \quad 
 \mbox{for} \quad \Lam \to 0
\end{equation}
in two dimensions. Consequently, also the radial mass $m_{\sg}^2$
of the Bose fields vanishes linearly in $\Lam$. 
In three dimensions one has 
$\int_{q|\Lam} G_{\pi}^2(q) \propto \Lam^{-1}$
for small $\Lam$ such that $\lam$ and $m_{\sg}^2$ scale to zero 
logarithmically for $\Lam \to 0$.
Since $m_{\sg}^2$ is the dominant contribution to the denominator
of $G_{\sg}$ at small momenta and frequencies, the scaling of
$m_{\sg}^2$ to zero as a function of $\Lam$ implies that $G_{\sg}$ 
(at $\Lam = 0$) diverges as
\begin{eqnarray}
 G_{\sg}(sq) \propto & s^{-1} & \mbox{for} \; d=2 \\
 G_{\sg}(sq) \propto & \log s & \mbox{for} \; d=3
\end{eqnarray}
in the limit $s \to 0$. 
Although derived from an approximate truncation of the functional
flow equation, this result is {\em exact} even in two dimensions,
where the renormalization of $m_{\sg}^2$ is very strong.
This is due to the fact that the scaling dimension of $m_{\sg}^2$ 
is fully determined by the scaling dimension of the Goldstone 
propagator and the existence of a fixed point for $\tilde\lam$, 
but does not depend on the position of the fixed point.
\cite{pistolesi04}
 
The flow of $Z_{\sg}$ is given by 
\begin{equation}
 \partial_{\Lam} Z_{\sg} = \frac{(\lam\alf)^2}{2}
 \int_{q|\Lam} \left. 
 \partial_{p_0}^2 G_{\pi}(p+q) G_{\pi}(q) \right|_{p=0} 
\end{equation}
for small $\Lam$. The integral over the second derivative of
$G_{\pi} G_{\pi}$ is of order $\Lam^{d-6}$.
In two dimensions the coupling $\lam$ vanishes linearly in
$\Lam$, such that $\partial_{\Lam} Z_{\sg} \propto \Lam^{-2}$,
implying that $Z_{\sg}$ diverges as $\Lam^{-1}$ for $\Lam \to 0$.
Hence, the term $Z_{\sg} q_0^2$ with $|q_0|=\Lam$ in the
denominator of $G_{\sg}$ scales linearly in $\Lam$, as $m_{\sg}^2$.
In three dimensions 
$\int_{q|\Lam} \partial_{p_0}^2 G_{\pi}(p+q) G_{\pi}(q) |_{p=0}$ 
diverges as $\Lam^{-3}$, while $\lam$ vanishes only logarithmically.
Hence $\partial_{\Lam} Z_{\sg} \propto (\log\Lam)^{-2} \Lam^{-3}$,
which is larger than in two dimensions. Integrating over $\Lam$
one finds $Z_{\sg} \propto (\Lam\log\Lam)^{-2}$, which means that
$Z_{\sg} q_0^2$ vanishes as $(\log\Lam)^{-2}$ in the infrared
limit. This yields a subleading logarithmic correction to the 
mass term $m_{\sg}^2$ in the denominator of $G_{\sg}$. An analogous 
analysis with the same results as just obtained for $Z_{\sigma}$ also holds 
for the momentum renormalization factor $A_{\sigma}$. 
A strong renormalization of longitudinal correlation functions 
due to Goldstone fluctuations appears in various physical contexts. 
\cite{weichman88,zwerger04}
Recently, a singular effect of Goldstone fluctuations on the
fermionic excitations in a superfluid was found in 
Gaussian approximation.\cite{lerch08} This singularity appears
only after analytic continuation to real frequencies, and its fate
beyond Gaussian approximation remains to be clarified.

Since $m_{\sg}^2$ and $\lam$ scale to zero in the infrared limit
in $d \leq 3$, all purely bosonic contributions to the effective
action scale to zero. On the other hand, the fermion-boson coupling 
remains finite.
One is thus running into a strong coupling problem, indicating a 
failure of our truncation, if fermionic fields are 
integrated too slowly, compared to the bosons.
The problem manifests itself particularly strikingly in the flow
equation for the order parameter, Eq.~(\ref{eq:alpha_ssb}), in 
two dimensions.
Since $m_{\sg}^2 \propto \Lam_b$ for small $\Lam_b$, the fermionic 
contribution to $\partial_{\Lam}\alf$ is of order $\Lam^{-1}$
if one chooses $\Lam_f = \Lam_b = \Lam$, leading to a spurious 
divergence of $\alf$ for $\Lam \to 0$.

The problem can be easily avoided by integrating the fermions
fast enough, choosing $\Lam_f \ll \Lam_b$ in the infrared limit.
In our numerical solution of the flow equations in the following
section we will choose $\Lam_b = \Lam$ and $\Lam_f = \Lam^2/\Lam_c$ 
for $\Lam < \Lam_c$, which matches continuously with the equal 
choice of cutoffs for $\Lam > \Lam_c$. 
The fermionic contribution to $\partial_{\Lam}\alf$ in 
Eq.~(\ref{eq:alpha_ssb}) is then finite for small $\Lam$, since
the factor $\Lam'_f = 2\Lam$ in $\int_{k|\Lam_f}$ compensates 
the divergence of $m_{\sg}^{-2}$ in front of the integral.
Since the fermions are gapped below $\Lam_c$, one could also
integrate them completely (set $\Lam_f$ to zero),
and then compute the flow driven by $\Lam_b$ only. 
The freedom to choose fermionic and bosonic cutoffs independently 
was exploited also in a recent fRG-based computation of the 
fermion-dimer scattering amplitude in vacuum.\cite{scherer07}

\subsection{Numerical results in two dimensions}
\begin{figure}
\begin{center}
\includegraphics*[width=75mm]{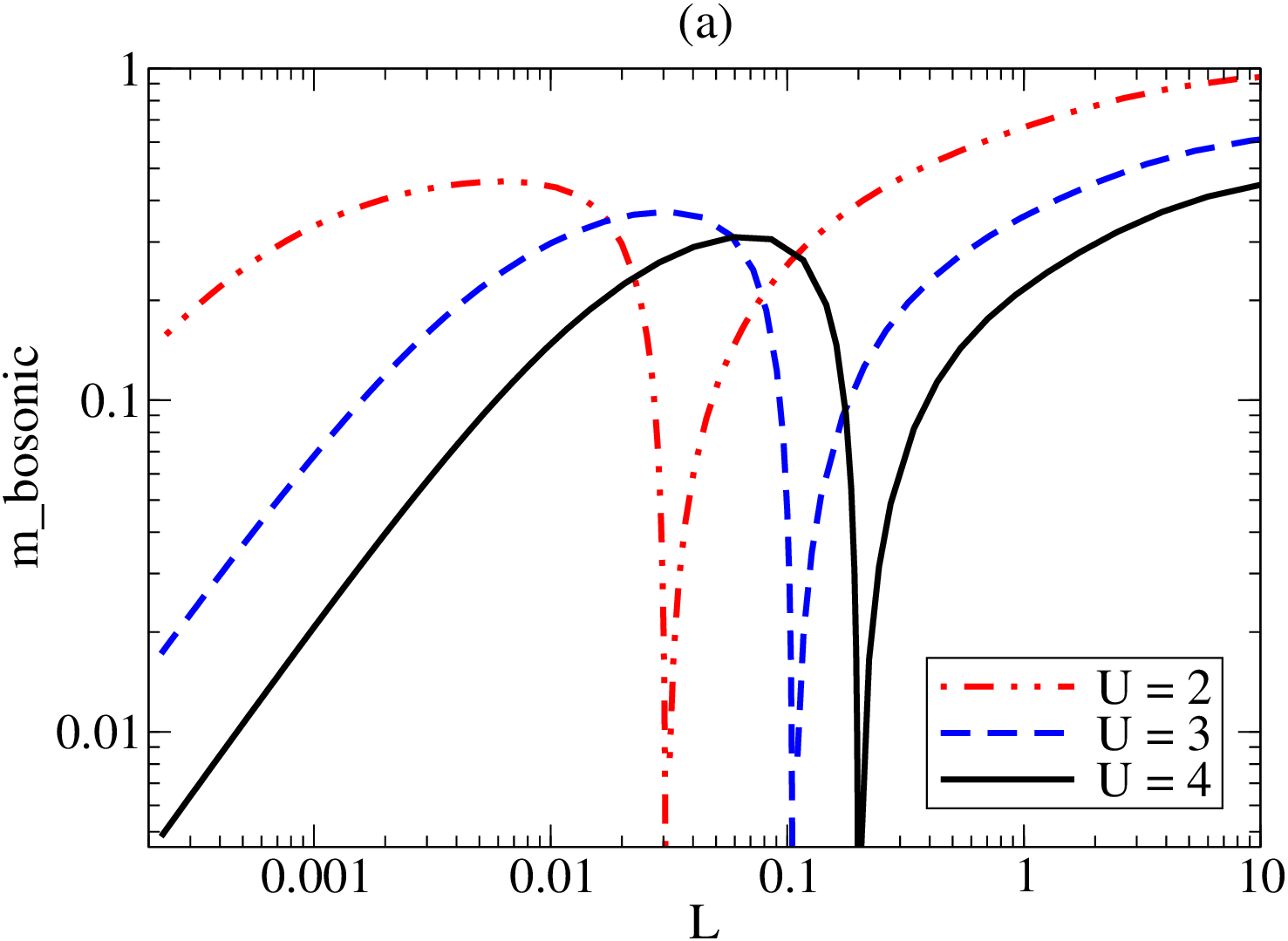}\\[7mm]
\hspace{-1mm}\includegraphics*[width=77mm]{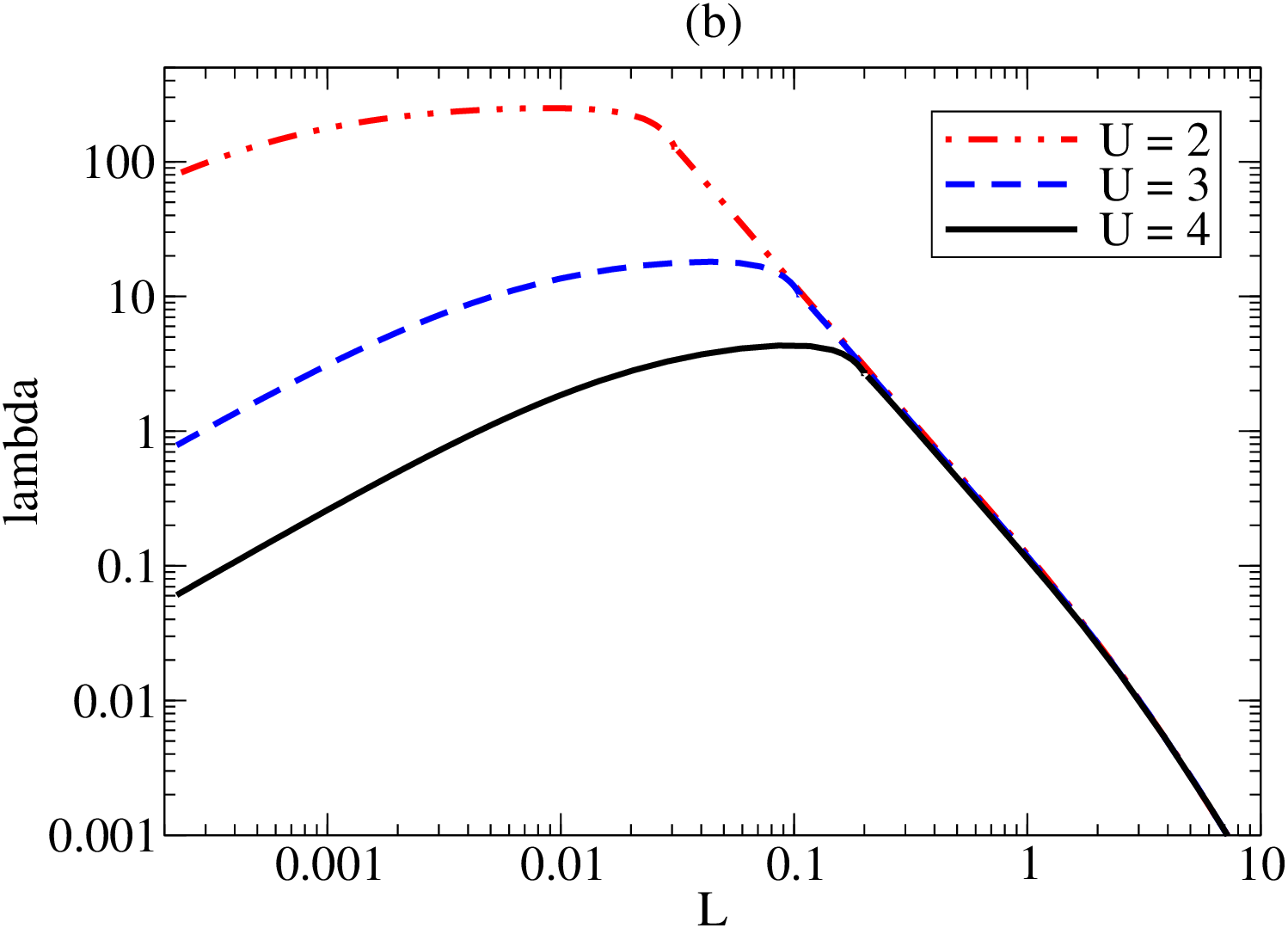}
\caption{(color online)
 (a): Flows of the bosonic mass, $m^{2}_{b}$ for 
 $\Lambda>\Lambda_{c}$ and $m^{2}_{\sigma}$ for 
 $\Lambda<\Lambda_{c}$. 
 (b): Quartic coupling $\lambda$.}
\label{fig:mSig}
\end{center}
\end{figure}
In this section, we present a numerical solution of our flow 
equations from Sec.~\ref{subsec:flow_equations} in two dimensions. 
Technically, we employ a fifth-order Runge-Kutta integration 
routine\cite{GSL} to solve coupled, ordinary differential equations.
At each increment of the Runge-Kutta routine, two-dimensional 
integrations over the whole Brillouin zone have to be executed. 
For this purpose, we employ an integrator for singular functions
\cite{GSL} with relative error of less then $1\%$. 
In particular, for $\Lam\approx\Lam_c$ it is imperative to 
operate with sufficiently accurate routines as the integrands 
are large and small deviations result in a significant spread in 
the final values for $\Lambda\rightarrow0$.

We fix our energy units by setting the hopping amplitude $t=1$.
We choose a chemical potential $\mu=-1.44$ corresponding to an 
average electron density of $1/2$ (quarter-filled band).
This choice represents the generic case of a convex Fermi surface
remote from van Hove singularities.
The only varying input parameter is the Hubbard U, which determines 
the initial value of the bosonic mass via $m^{2}_b = |2/U|$.

Initially, the flow starts in the symmetric regime with 
$\Lam = \Lam_0 = 100$, where Eqs.~(\ref{eq:mass_sym} - 
\ref{eq:lambda}) determine the evolution. 
The critical scale is determined by the condition 
$m_{b}^{2}(\Lambda_c)=0$. 
In the symmetry-broken regime ($\Lambda<\Lambda_{c}$), 
Eqs.~(\ref{eq:alpha_ssb} - \ref{eq:g_pi_ssb}) determine the 
evolution.

\begin{figure}
\begin{center}
\includegraphics*[width=73mm]{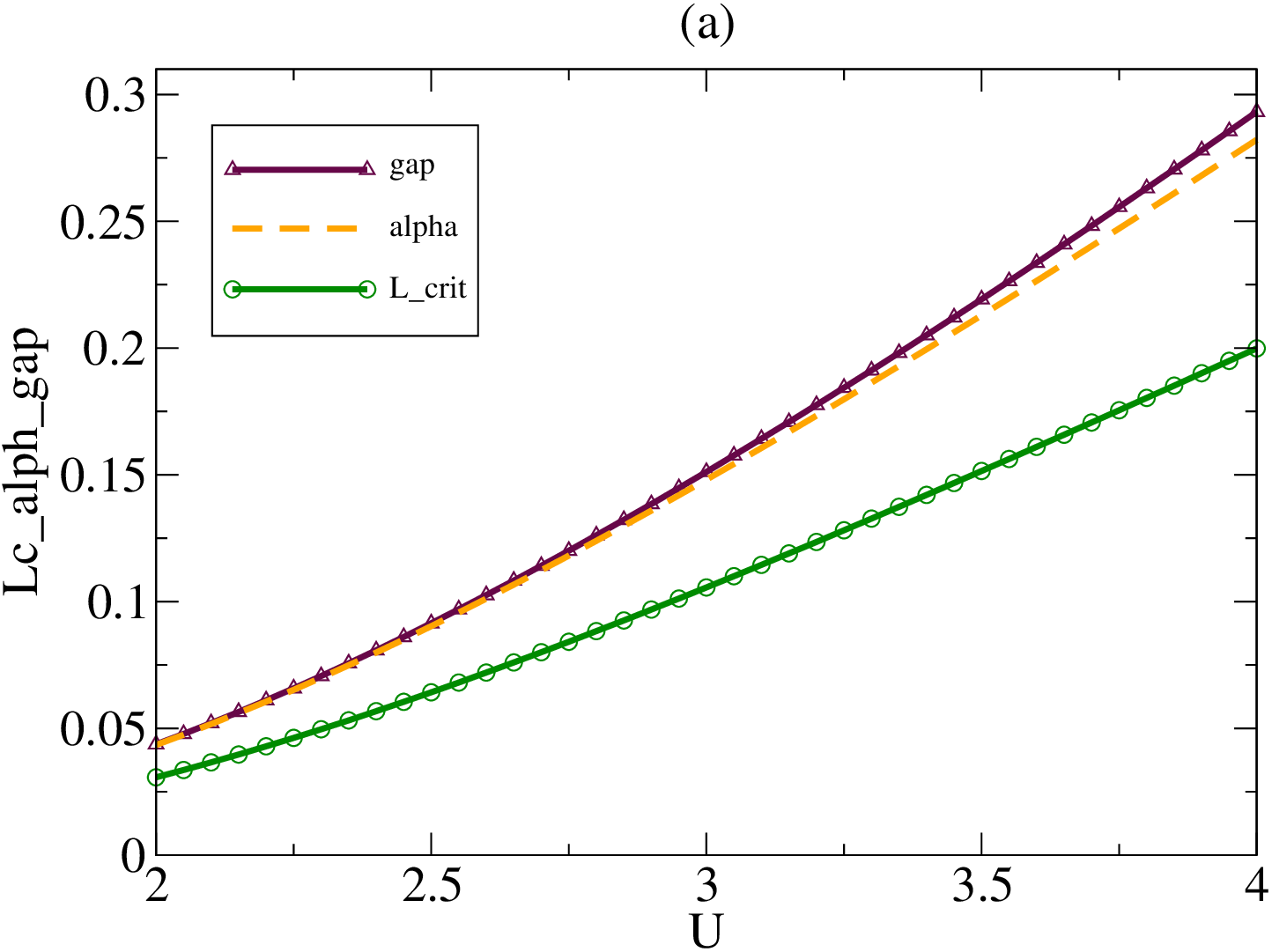}\\[7mm]
\includegraphics*[width=73mm]{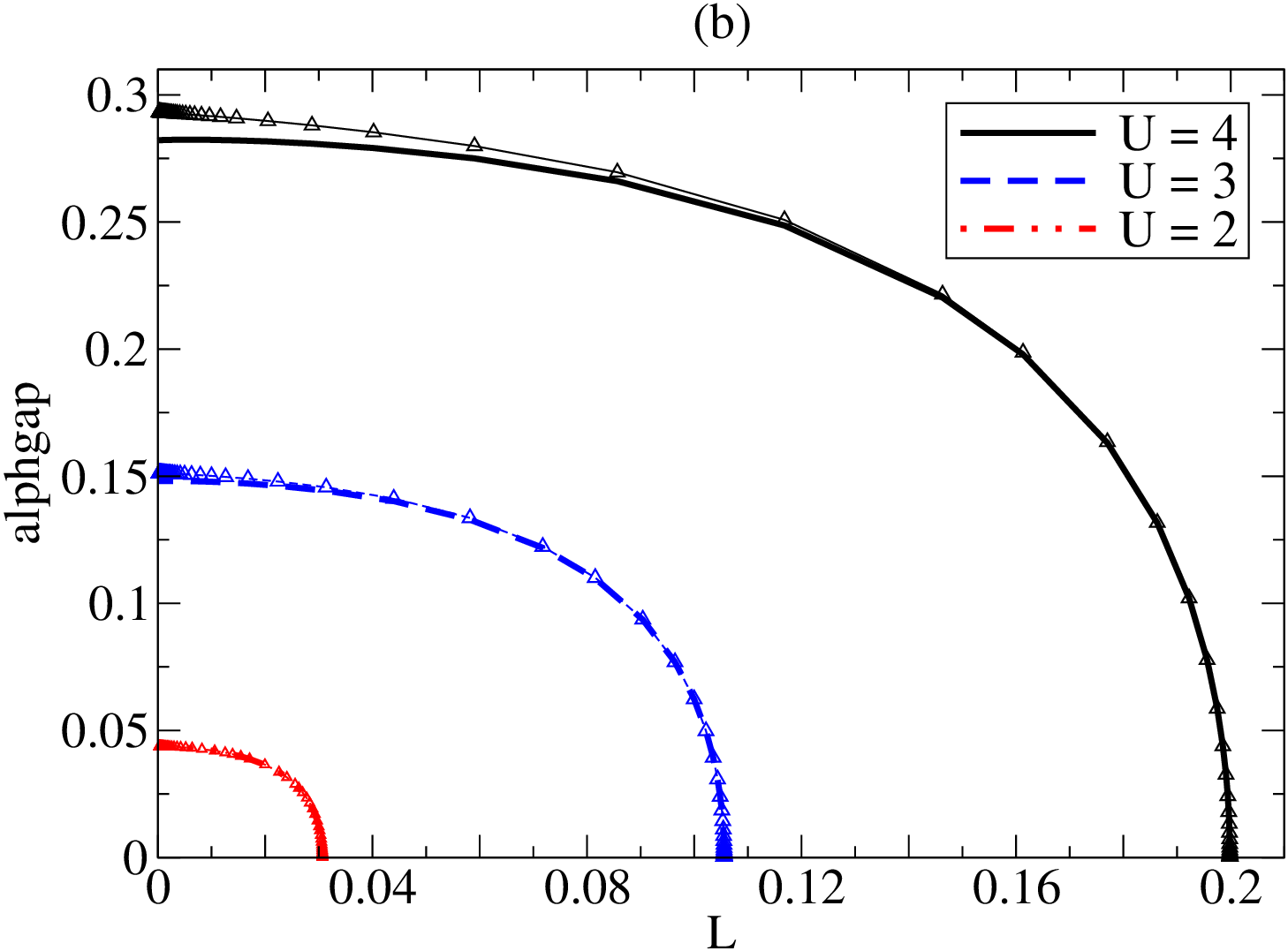}
\caption{(color online)
 (a): Fermion gap $\Delta$, order parameter $\alpha$, and the 
 critical scale $\Lambda_{c}$ versus Hubbard $U$. 
 (b): Exemplary flows for $\Delta$ (triangles) and $\alpha$ (lines) 
 each corresponding to one point in (a).}
\label{fig:gap}
\end{center}
\end{figure}
In Fig.~\ref{fig:mSig} characteristic flows of the bosonic mass 
and the quartic coupling are shown in double-logarithmic plots for 
different choices of the Hubbard U. The sharp de- and increase of 
the bosonic mass marks the region around $\Lam_c$. 
For small $\Lam$, the flow reaches the infrared asymptotic regime 
(see Sec.~\ref{subsec:asymptotics}). 
The scale $\Lambda_{\text{IR}}$ at which this scaling sets in 
decreases for decreasing $U$. The numerically obtained scaling
$m^{2}_{\sigma},\,\lambda \propto \Lam$ is consistent with the 
analytical result of Sec.~\ref{subsec:asymptotics}.

In Fig.~\ref{fig:gap} (a) we compare the fermion gap with the 
bosonic order parameter (final values at $\Lam = 0$) and the 
critical scale as a function of $U$. 
In Fig.~\ref{fig:gap} (b) the flow of $\Delta$ and $\alf$ as
a function of $\Lam$ is shown for various choices of $U$.
We observe $\Delta,\,\alpha \propto (\Lam_c - \Lam)^{1/2}$ for 
$\Lambda\lesssim\Lambda_{c}$ as derived below Eq.~(\ref{eq:alpha_L_c}).
The ratio $\Delta/\Lam_c$, where $\Delta$ is the final gap for
$\Lam \to 0$, is approximately 1.4 for the values of $U$ studied
here.
As a result of fluctuations, the gap is reduced considerably 
compared to the mean-field result
\begin{equation}
 \frac{\Delta}{\Delta_{\rm BCS}} \approx 0.25
 \;
\end{equation}
for $2 \leq U \leq 4$.
The main reduction here stems from the bosonic self-interactions 
in the symmetric regime leading to a substantial decrease of 
$\Lam_{c}$ via the second term of Eq.~(\ref{eq:mass_sym}).
A reduction of the gap compared to the mean-field value is
generally present even in the weak coupling limit $U \to 0$.
In fermionic perturbation theory second order corrections reduce 
the prefactor of the BCS gap formula even for $U \to 0$.\cite{martin92}
The reduction obtained here is slightly stronger than what is
expected from a fermionic renormalization group calculation.
\cite{gersch08}

For $\Lam < \Lam_c$, Goldstone fluctuations slightly reduce $\alf$ 
via the term involving $G_{\pi}(q)$ in Eq.~(\ref{eq:alpha_ssb}).
On the other hand, the term due to the Goldstone mode
in Eq.~(\ref{eq:gap_ssb}) enhances the fermionic gap $\Delta$
relative to $\alf$, such that $\Delta$ is generally slightly
larger than $\alf$. This difference will become larger upon increasing 
the interaction strength as one enters the regime of a Bose gas made of tightly bound fermions. 
Here, however, for relatively weak interactions, the impact of Goldstone fluctuations
on both $\alf$ and $\Delta$ is very modest.
By contrast, the impact of Goldstone fluctuations is known to be 
dramatic at finite temperatures (not treated here) in two dimensions,
since they drive the order parameter to zero.\cite{goldenfeld_book} 

\begin{figure}
\begin{center}
\includegraphics*[width=78mm]{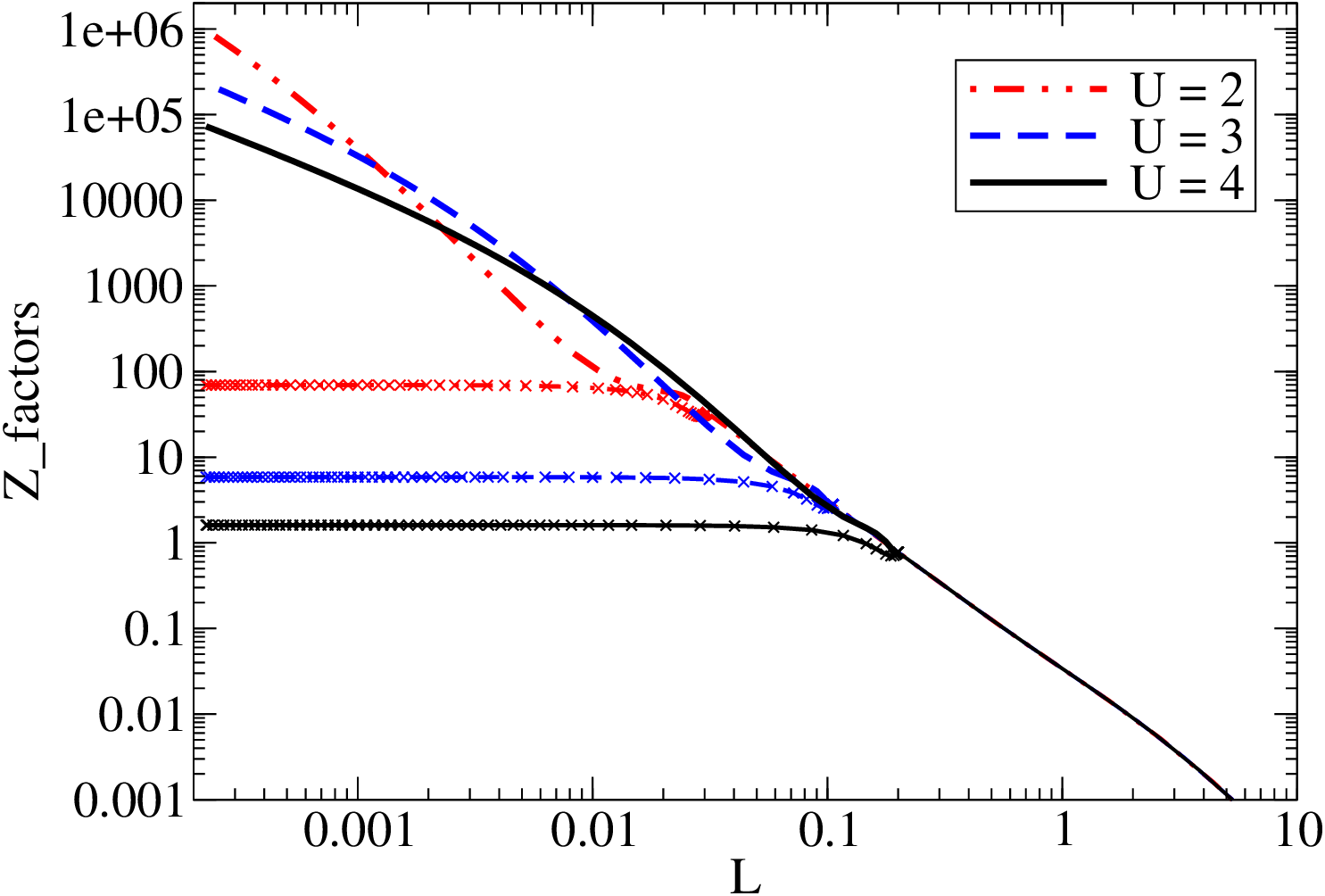}
\caption{(color online)
 Flows of $Z_{\sigma}$ (lines) and $Z_{\pi}$ (crosses).}
\label{fig:Z_factors}
\end{center}
\end{figure}
\hspace{5mm}
\begin{figure}
\begin{center}
\includegraphics*[width=75mm]{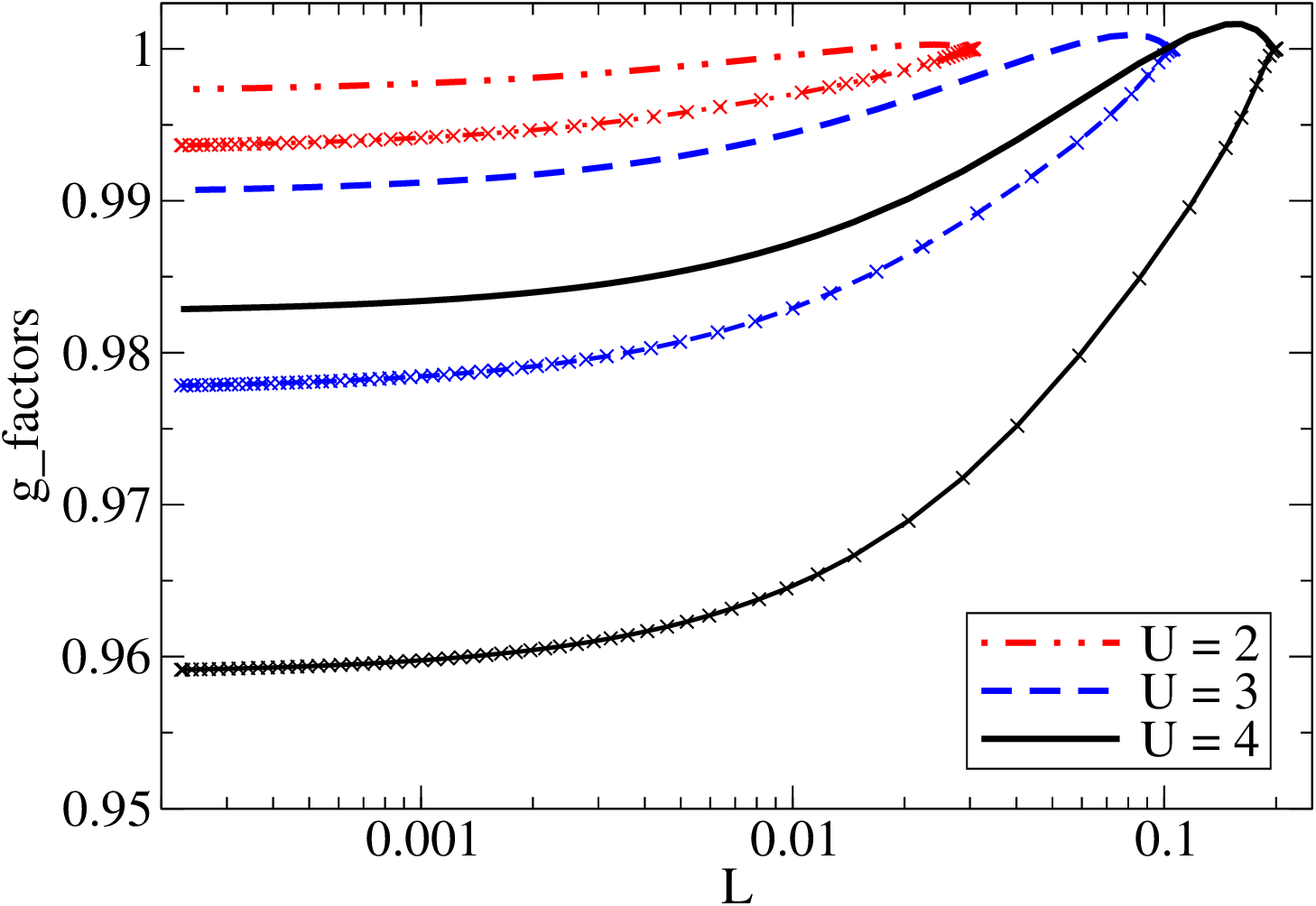}
\caption{(color online) 
 Flows of fermion-boson vertices, $g_{\sigma}$ (lines), 
 and $g_{\pi}$ (lines with crosses), for $\Lam < \Lam_c$.}
\label{fig:vertex}
\end{center}
\end{figure}
In Fig.~\ref{fig:Z_factors}, we show flows of the Z-factors 
of the $\sigma$- and $\pi$-field, respectively. 
In the symmetric regime, the evolution is independent of $U$. 
At $\Lam_{c}$, the $\phi$-field splits into the $\sigma$- and 
$\pi$-modes with $Z_{\sigma}$ diverging in the limit $\Lam \to 0$ 
as $Z_{\sigma} \propto \Lam^{-1}$ 
(cf.\ Sec.~\ref{subsec:asymptotics}). 
The Z-factor of the Goldstone field saturates for 
$\Lam \ll \Delta$. 
The flows for the A-factors (not shown) parametrizing the 
momentum dependence of the $\sigma$- and $\pi$-propagators 
exhibit very similar behavior.

Finally, in Fig.~\ref{fig:vertex} we show flows of the 
fermion-boson vertices $g_{\sigma}$ and $g_{\pi}$ for $\Lam < \Lam_c$. 
Their relative changes are only of the order of a few percent 
(note the scale of the vertical axis) with $g_{\sigma}$ being a 
bit larger than $g_{\pi}$.

\section{Conclusion}

Truncating the exact fRG flow, we have derived approximate flow
equations which capture the non-trivial order parameter fluctuations
in the superfluid ground state of the attractive Hubbard model,
which has been chosen as a prototype model for attractively 
interacting fermions.
The superfluid order parameter is associated with a bosonic field
which is introduced via a Hubbard-Stratonovich decoupling of the
fermionic interaction.
Below a critical scale $\Lam_c$, the bosonic effective potential
assumes a mexican hat shape leading to spontaneous symmetry
breaking and a Goldstone mode. The bosonic order parameter is
linked to but not equivalent to a fermionic gap.
The fermionic gap is significantly smaller than the mean-field gap,
mostly due to fluctuations above the scale $\Lam_c$.
Transverse order parameter fluctuations (Goldstone mode) below
$\Lam_c$ lead to a strong renormalization of radial fluctuations.
The radial mass and the bosonic self-interaction vanish linearly
as a function of the scale in two dimensions, and logarithmically 
in three dimensions, in agreement with the exact behavior of an 
interacting Bose gas.\cite{pistolesi04}
On the other hand, the average order parameter, the fermionic gap, 
and the interaction between fermions and bosons are affected only
very weakly by the Goldstone mode.

Supplementing the flow equations derived above by a shift of the
chemical potential, to keep the density fixed, one may also try
to deal with larger values of $U$. 
Eagles \cite{eagles69} and Leggett \cite{leggett80} have shown
that already the BCS mean-field theory captures many features of
the condensed Bose gas ground state made from strongly bound
fermion pairs in the limit of strong attraction.
Beyond mean-field theory, the difference between the fermionic
gap and the order parameter $\alpha$ increases at larger $U$.
It will also be interesting to extend the present analysis to $T>0$,
in particular in view of the possibility of a finite fermionic gap 
in the absence of long-range order in a Kosterlitz-Thouless phase 
at low finite temperatures.




\begin{acknowledgments}
We thank S. Diehl, S. Floerchinger, H. Gies, P. Jakubczyk, A. Katanin, H. C. Krahl, 
P. Kopietz, J. Pawlowski, M. Salmhofer, M. Scherer, and R. Zeyher for very useful discussions. 
S. Diehl and P. Jakubczyk are gratefully 
acknowledged for critically reading the manuscript.
\end{acknowledgments}

\end{document}